\RequirePackage{xcolor}
\documentclass{IEEEtran}
\usepackage{cite}
\usepackage{amsmath,amssymb,amsfonts}
\usepackage{hyperref}
\usepackage{cleveref}
\usepackage[noend]{algpseudocode}
\usepackage{graphicx}
\usepackage{textcomp}
\usepackage{booktabs}
\usepackage{bbm}
\usepackage{fontawesome}
\usepackage{mathtools}
\usepackage{array}
\usepackage{booktabs}
\usepackage{multirow}
\usepackage{siunitx}
\usepackage{xpatch}

\makeatletter
\xpatchcmd{\algorithmic}
  {\ALG@tlm\z@}{\leftmargin\z@\ALG@tlm\z@}
  {}{}
\makeatother

\newcommand{\bftab}{\fontseries{b}\selectfont}
\newcommand{\PreserveBackslash}[1]{\let\temp=\\#1\let\\=\temp}
\newcolumntype{L}[1]{>{\raggedright\arraybackslash}m{#1}}
\newcolumntype{C}[1]{>{\centering\arraybackslash}m{#1}}
\newcolumntype{R}[1]{>{\raggedleft\arraybackslash}m{#1}}

\def\BibTeX{{\rm B\kern-.05em{\sc i\kern-.025em b}\kern-.08em
    T\kern-.1667em\lower.7ex\hbox{E}\kern-.125emX}}
\begin{document}
\bstctlcite{IEEEexample:BSTcontrol}

\title{Deep Learning-based Conditional Inpainting for Restoration of Artifact-affected 4D CT Images}
\author{F. Madesta, T. Sentker, T. Gauer, and R. Werner
\thanks{This work was funded by DFG research grants WE 6197/2-1 and WE 6197/2-2 (project number 390567362). RW and TG were further supported by a research grant from Siemens Healthineers AG.}
\thanks{F. Madesta, T. Sentker and R. Werner are with the Department of Computational Neuroscience and the Center for Biomedical Artificial Intelligence (bAIome), University Medical Center Hamburg-Eppendorf, 20246 Hamburg, Germany (e-mail: \{f.madesta, t.sentker, r.werner\}@uke.de)}
\thanks{T. Gauer is with the Department of Radiotherapy and Radio-Oncology, University Medical Center Hamburg-Eppendorf, 20246 Hamburg, Germany (e-mail: t.gauer@uke.de).}
}
\maketitle

\begin{abstract}

%%%%%%%%%%%% --- 250 WORDS MAX -> 

4D CT imaging is an essential component of radiotherapy of thoracic/abdominal tumors. 
4D CT images are, however, often affected by artifacts that compromise treatment planning quality. In this work, deep learning (DL)-based conditional inpainting is proposed to restore anatomically correct image information of artifact-affected areas.     
The restoration approach consists of a two-stage process: DL-based detection of common interpolation (INT) and double structure (DS) artifacts, followed by conditional inpainting applied to the artifact areas. In this context, \emph{conditional} refers to a guidance of the inpainting process by patient-specific image data to ensure anatomically reliable results. 
The study is based on 65 in-house 4D CT \textcolor{black}{images} of lung cancer patients (48 with only slight artifacts, 17 with pronounced artifacts) 
\textcolor{black}{and two publicly available 4D CT data sets that serve as independent external test sets.}   
Automated artifact detection revealed a ROC-AUC of 0.99 for INT and of 0.97 for DS artifacts (in-house data). 
The proposed inpainting method decreased the average root mean squared error (RMSE) by \textcolor{black}{52}\,\% (INT) and \textcolor{black}{59}\,\% (DS) for the in-house data.
\textcolor{black}{For the external test data sets, the RMSE improvement is similar (50\,\% and 59\,\%, respectively).} 
\textcolor{black}{Applied to 4D CT data with pronounced artifacts (not part of the training set), \textcolor{black}{72\,\%} of the detectable artifacts were removed.}	
The results highlight the potential of DL-based inpainting for restoration of artifact-affected 4D CT data. \textcolor{black}{Compared to recent 4D CT inpainting and restoration approaches}, the proposed methodology illustrates the advantages of exploiting patient-specific prior image information.
\end{abstract}


\section{Introduction}
\label{Sec:1}

4D CT imaging is a technological key component of radiotherapy (RT) of tumors undergoing respiratory motion, i.e., thoracic and abdominal tumors \cite{schmitt:2020}.  
A 4D CT data set consists of a series of 3D CT images that represents the patient's anatomy at different breathing phases and typically includes 10 phases. 

Common clinical 4D CT imaging protocols are based on simultaneous acquisition of CT projection data and a breathing signal. After CT scanning, the breathing signal information is used to assign respiratory phase information to the acquired CT projection data. The phase-labeled projection data is then used to reconstruct 3D CT images at the desired breathing phases. However, current CT scanners do usually not allow coverage of the entire thorax or abdomen within a single gantry rotation. Data acquisition therefore encompasses multiple breathing cycles of the patient that correspond to projection data acquired at different couch positions ($z$-positions) of the desired field-of-view. As a consequence, changes of the patient's breathing patterns and, in particular, pronounced breathing irregularities during data acquisition lead to artifacts in the reconstructed 3D CT phase images. 

Related artifacts in clinical 4D CT data occur frequently \cite{yamamoto:2008,werner:2017}, have been shown to compromise RT target volume delineation \cite{persson:2010,keall:2006} as well as dose calculation \cite{keall:2006,sothmann:2018}, and have been reported to be correlated with unfavorable patient outcome \cite{sentker:2020}. 
To reduce the number and amount of 4D CT artifacts, existing 4D CT scanning protocols have been optimized and novel 4D CT approaches proposed during the past two decades \cite{langner:2008,keall:2007,thomas:2014,pan:2017,werner:2019}. 
Despite progress in this field, even recently introduced commercial 4D CT scan modes do \emph{not} generate artifact-free 4D CT data sets \cite{werner:2021,szkitsak:2021}. This stresses the need to develop reliable post processing-based approaches to reduce and ideally remove the artifacts. 

\begin{figure*}
    \centering
    \includegraphics[]{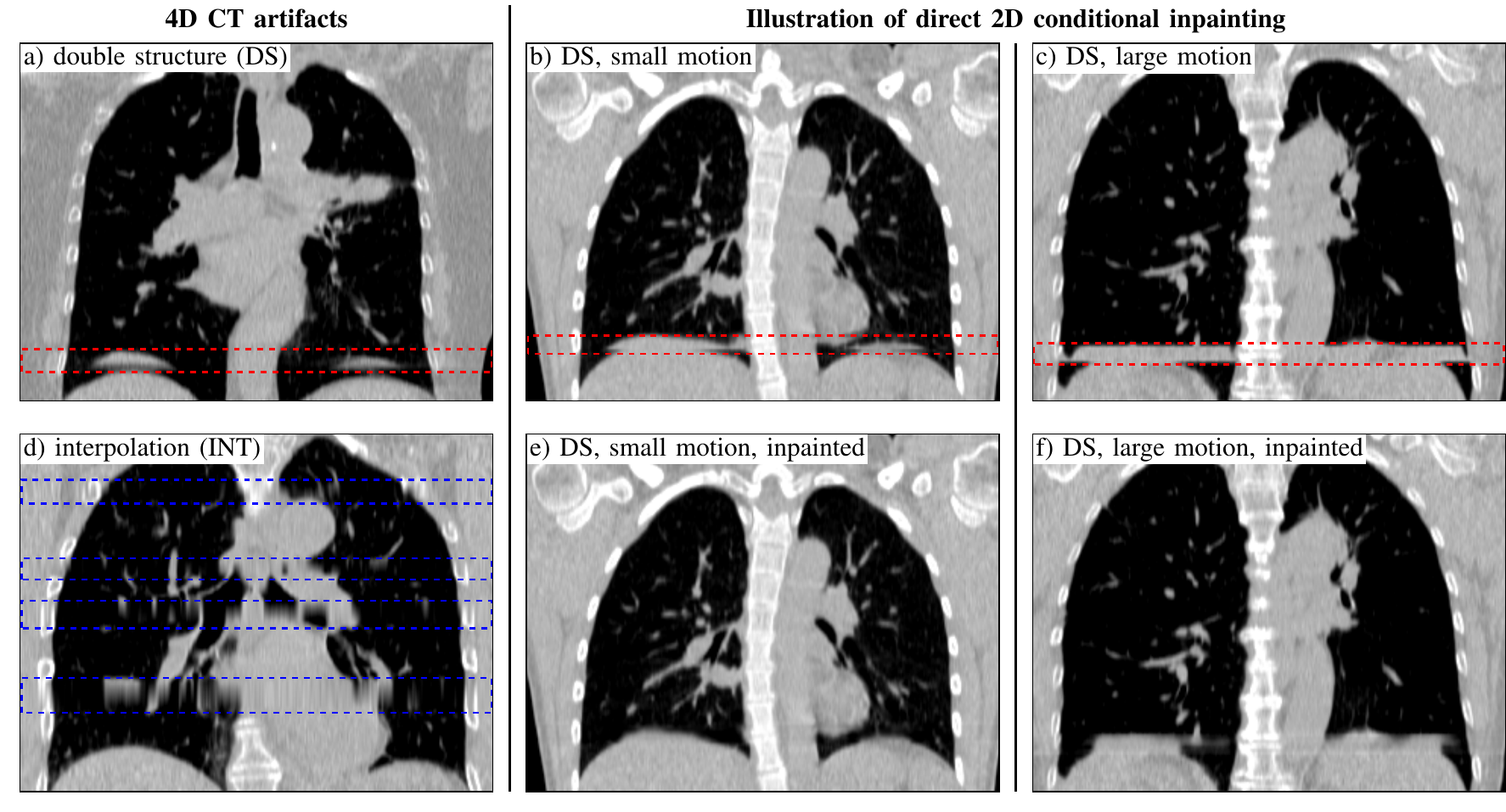}
    \caption{Left: Visualization of double structure (DS) artifacts (\textcolor{black}{subfigure a, artifact area highlighted by red box}) and interpolation (INT) artifacts (d, \textcolor{black}{artifact areas highlighted by blue boxes}) in 4D CT image data. 
    Right: Illustration of direct 2D conditional inpainting on 4D CT data affected by DS artifacts. \textcolor{black}{Subfigure b shows an artifact (highlighted in red) for a patient with a \emph{small} breathing amplitude and subfigure e the same coronal slice of the 3D CT volume after 2D inpainting. In c, a DS artifact (highlighted in red) is depicted for a patient with \emph{large} breathing amplitudes; the corresponding coronal slice after 2D inpainting is shown in f.} As visible, a 2D-based approach is able to properly inpaint smaller artifacts; however, it fails for more pronounced patient motion and artifacts. INT artifacts have not yet been addressed by inpainting approaches.}
    \label{fig:fig_1}
\end{figure*}

4D CT artifacts can be divided into two types: double structure (DS) and interpolation (INT) artifacts \cite{sentker:2020} (see \cref{fig:fig_1}\,a and d). DS artifacts (also referred to as duplication, overlapping, or misalignment artifacts \cite{yamamoto:2008,shao:2021}) are caused by different respiratory patterns, for instance due to breathing amplitude variability, during data acquisition at adjacent $z$-positions. The differences result in inconsistent breathing phase representations, leading to different appearances of anatomical structures at respective breathing phases and $z$-positions. In contrast, INT artifacts (also known as missing data artifacts \cite{shao:2021}) are due to a lack of projection data for reconstruction of image slices for the desired breathing phases and $z$-positions. The lack is usually caused by breathing pauses or suboptimal 4D CT protocol selection (mismatch of pitch factor and breathing period). Instead of, e.g., providing black areas for the affected slices, a common vendor-specific `solution' is to interpolate Hounsfield unit (HU) voxel values for the affected CT slices based on reconstructed neighboring $z$-slices.

Existing conventional image post-processing approaches to reduce DS artifacts cover reg\-is\-tra\-tion-based spatio-temporal image interpolation and sorting to optimize spatial-temporal structure smoothness \cite{ehrhardt:2007,he:2013}, graph-based structure alignment after reconstruction \cite{han:2011}, or geodesic density regression (GDR) to correct for artifacts by using corresponding regions of other (artifact-free) breathing phases \cite{shao:2021}. In particular the registration-based approaches are, however, usually associated with long run times, making an integration into clinical workflows difficult. Moreover, INT artifacts have not yet been addressed by these approaches.

In principle, the appearance and cause of INT artifacts -- missing image information for a limited region of the image -- suggest tackling these artifacts by image inpainting approaches.   
Motivated by the general success of deep learning (DL), in the natural image domain, especially generative adversarial network (GAN)-based inpainting approaches became competitive and showed promising results \cite{liu:2018,zhang:2019}. Consequently, these approaches were adapted to facilitate a filling of missing or corrupted information in medical images. Examples are the use of conditional GANs for inpainting of arbitrary shaped regions in 2D MRI slices \cite{armanious:2020} and inpainting of missing areas in 2D ultrasound images \cite{yu:2020}.
However, the inability of GANs to generalize and generate inpainting content with reasonable structure/visual reality is extremely problematic when it comes to medical images, where retaining information of minor abnormal changes is essential when a diagnoses is to be made \cite{yi:2019}. 

For 3D or even spatio-temporal medical image data, DL-based inpainting methods can only be sparsely found in the literature, potentially because GPU memory is often a limiting factor \cite{mori:2019} and approaches to efficiently handle the, in comparison to standard natural image data, large input data are required. 
\textcolor{black}{Jin \textit{et al.}, for instance,  proposed a 3D GAN-based inpainting approach to generate realistically appearing synthetic tumors in thoracic and abdominal CT images. The GAN was trained on and applied to 3D CT patches \cite{jin:2021}. 
Moreover,} Kang \textit{et al.} developed a patch-based approach to generate inpainted 3D brain MR images from intermittently sampled 2D images \cite{kang:2021}. 
However, due to potential patch-boundary artifacts, in thoracic volumetric image data, slice-wise approaches are often the preferred strategy \cite{madesta:2018}. 
Consistent with this, Mori \textit{et al.} provided an early slice-wise proof-of-concept that DL-based inpainting can help to reduce 4D CT artifacts \cite{mori:2019}. Focusing again only on DS artifacts, they trained a straight-forward autoencoder to `translate' artifact-affected into artifact-free coronal and sagittal CT slices. Although working only on slices, i.e., on 2D data, the CT input data had to be down-sampled and down-sized to match GPU memory requirements. The authors showed their approach to be promising, but evaluation data was neither shown for 3D data nor for real artifact-affected 4D CT data. 
Solely working on 2D coronal or sagittal CT slices may indeed be sufficient for small breathing motion amplitudes; however, for real patient data and in particular for patients with larger motion amplitudes, anatomical structures often do not remain within a single slice for all breathing phases. As a consequence, stacking post-processed slices obtained by direct 2D inpainting leads to inconsistent structure appearance in the resulting volumetric CT data and insufficient artifact reduction.
\textcolor{black}{This is illustrated in \cref{fig:fig_1}: Fig.~\ref{fig:fig_1}\,b shows a small DS artifact for a patient with small breathing motion amplitudes and \cref{fig:fig_1}\,c a pronounced DS artifact for a patient with large motion amplitudes. While 2D inpainting is sufficient to almost remove the artifact in the first case (\cref{fig:fig_1}\,e), it fails in the second case (\cref{fig:fig_1}\,f).} 
Moreover and problematically, the \textcolor{black}{straightforward inpainting approaches like the} method proposed by Mori \textit{et al.} also lead to undesired significant alteration of the structures in the \emph{artifact-free} areas, that is, consistency of the originally acquired and the post-processed images is not guaranteed for the image regions not affected by the artifacts. 
The present work is motivated by the early proof-of-concept \textcolor{black}{of Mori \textit{et al.}} \cite{mori:2019}, but presents solutions to the shortcomings from the methodical as well as the evaluation perspective. 

Thus, we propose a DL-based conditional inpainting approach for 4D CT data that automatically detects and fills image areas affected by motion artifacts. The particular foci were (i) anatomically correct inpainting and (ii) clinical applicability. To achieve (i), we extend the idea of straight-forward to conditional inpainting, making use of patient-specific prior image information. 
Furthermore, the proposed network operates on three-dimensional data 
and combines the concepts of partial convolutions \cite{liu:2018} (filling missing information from the outside in) and spatial transformers \cite{jaderberg2015} (guiding the network in the process of finding spatial correspondence) to maximize anatomically correctness and to be able to handle large motion amplitudes. To demonstrate (ii), we apply and evaluate respective developments for clinical in-house as well as 
\textcolor{black}{for two publicly available 4D CT data sets that serve as independent external test data}.

% ---------- MATERIALS SECTION:

\section{Materials and Methods}
\label{Sec:2}

To foster re-use and to ensure reproducibility, the trained DL models are publicly available at \url{github.com/IPMI-ICNS-UKE/[available-after-accept]}

\subsection{4D CT data}
\label{Sec:data}
The experiments were based on anonymized planning 4D CT data sets of 65 lung tumor patients treated with stereotactic body RT at our clinic (in-house data set; retrospective data evaluation approved by the local ethics board and the need to obtain written informed consent waived [WF-82/18]) and, \textcolor{black}{as independent external evaluation cohorts, the 10 thoracic 4D CT images of the open data repository DIR-Lab \cite{castillo2009a,castillo2009b} and 20 4D CT images of locally-advanced, non-small cell lung cancer patients of the 4D-Lung data repository \cite{balik2013,hugo2016,hugo2017}.} 

The in-house 4D CT images were acquired during free breathing (retrospective reconstruction; spiral scanning with 0.5\,s gantry rotation time, pitch factor  0.09), reconstructed as 10-phase 4D CT images and a corresponding temporal average 3D CT image (image spatial resolution: \mbox{0.98\,mm $\times$ 0.98\,mm $\times$ 2\,mm}, image dimensions in voxels: 512 $\times$ 512 $\times$ [150 -- 160]; data acquisition with Siemens Definition AS Open, Siemens Healthineers, Germany). Breathing curves for 4D CT reconstruction were acquired by the Varian RPM system (Varian Medical Systems, USA). At our clinic, both the phase images and the temporal average CT image are routinely used for target volume and organs at risk delineation (phase images and average CT) and dose calculation (average CT).  

The in-house 4D CT data sets were split into three groups by two imaging experts ($>$5 years of experience), according to the artifact severity visible in coronal and sagittal slices of the phase images: \emph{no artifacts group}, 21 data sets; \emph{slight artifacts group,} 27 data sets; and \emph{pronounced artifacts group}, 17 data sets. 
If at least one phase image of a 4D CT data set contained a pronounced artifact (independent of the artifact type), it was grouped into the pronounced artifacts group. 

The DIR-Lab data consist of 10 artifact-free 4D CT lung data sets provided by the University of Texas M.D. Anderson Cancer Centers (Houston, USA). Each 4D CT comprises 10 3D CT phase images (spatial resolution: [0.98 -- 1.16]\,mm $\times$ [0.98 -- 1.16]\,mm $\times$ 2.5\,mm, image dimensions: [256, 512] $\times$ [256, 512] $\times$ [94 -- 136]). Corresponding temporal average CT were computed based on the phase images.

\textcolor{black}{The 4D-Lung data set comprises 4D CT images of 20 non-small cell lung cancer patients and is publicly provided as part of the Cancer Imaging Archive. The CT image data were acquired with a 16-slice helical CT scanner (Philips Medical Systems). 
Each 4D CT comprises 10 3D CT phase images (spatial resolution: 0.98\,mm $\times$ 0.98\,mm $\times$ 3\,mm, image dimensions: 512 $\times$ 512 $\times$ [77 -- 149]). Temporal average CT images were again computed based on the different phase images. Two of the 20 4D CT images could not be read (missing DICOM files), resulting in 18 4D CT data sets, which were visually inspected and split into the above artifact groups (no artifacts: 3, slight artifacts: 5, pronounced artifacts: 10 cases). The data contained only DS artifacts.}

For all CT images, lung segmentations and bounding boxes were computed by an U-Net trained/tested on 666/222 images of the LUNA16 data set \cite{LUNA16} (Dice coefficient, test set: 0.99).

\subsection{Simulation of 4D CT artifacts}
\label{Sec:AF_simulation}
To obtain corresponding artifact-affected/artifact-free image pairs $\mathbf{I}_{\mathrm{A},i}$ and $ \mathbf{I}_i$, both of 
size $n_x\times n_y\times n_z$ and at a specific breathing phase $i\in\{1,\dots,n_{\text{ph}}\}$ of the very same 4D CT image $\left(\mathbf{I}_i\right)_{i\in\{1,\dots,n_\text{ph}\}}$ with $n_{\text{ph}}$ breathing phases, a realistic simulation of interpolation (INT) and double structure (DS) artifacts was required. 

For the Siemens scanner and software used for acquisition and reconstruction of the in-house data, INT artifacts are due to a violation of the data sufficiency condition (DSC) \cite{pan:2005}. The DSC describes a condition on the maximum breathing period of a patient during scanning for a given pitch factor during spiral 4D CT imaging (and vice versa). In the case of a DSC violation, the patient breathes too slowly at least during a specific scanning period to be able to reconstruct the corresponding axial CT image slices for all desired breathing phases, given the acquired projection data. This absence of suitable projection data is compensated during reconstruction via approximately linear interpolation of the HU values of the desired axial image slices between validly reconstructed axial image slices (\cite{werner_technical_2018}; actual vendor implementation not accessible by the authors). Thus, INT artifacts were simulated by voxel-wise 1D linear interpolation between HU values of two defined axial image slices. Therefore, the interpolated image slices do not carry any valid anatomical information.

DS artifacts result from changes of the patient breathing patterns between data acquisition at adjacent couch positions. The changes lead to different appearance of internal structures for similar breathing phases when the latter are (as it is usually the case) solely defined based on an external breathing signal, that is, they represent different physiological states. During reconstruction, the different physiological states manifest in visible discontinuities of organ/structure borders between axial slides of the affected adjacent couch positions of the phase CT images. 
To simulate DS artifacts in a defined $z$-range, we (1) defined the desired range by means of a binary mask $\mathbf{M}\in\{0,1\}^{n_x\times n_y\times n_z}$ that contains 1's for voxels of slices that represented the simulated artifact area, (2) applied a Gaussian filtering to $\mathbf{M}$ to obtain a smooth version $\tilde{\mathbf{M}}\in\mathbb{R}^{n_x\times n_y\times n_z}$ of $\mathbf{M}$, and (3) computed a DS artifact-affected version $\mathbf{I}_{\textbf{A},i}$ of the artifact-free basis image $\mathbf{I}_i$ by 
\begin{equation}
\label{eq:AF_simulation}
\mathbf{I}_{\textbf{A},i}=(\mathbf{1}-\tilde{\mathbf{M}})\odot \mathbf{I}_i+\tilde{\mathbf{M}}\odot \mathbf{I}_{j\neq i}
\end{equation}
with $\mathbf{1}$ as a matrix with the size of $\mathbf{M}$ that only contains 1's and $\odot$ as element-wise multiplication. That is, in the defined artifact area, the image information of a different phase image $\mathbf{I}_{j\neq i}$ of the considered 4D CT image was inserted, leading to typical DS artifacts and structure border inconsistencies. Step (2) was motivated by a blurred transition between artifact-free and -affected areas often visible in clinical 4D CT data sets.
For DS artifacts, $\mathbf{I}_{\textbf{A},i}$ phase images were simulated for all breathing phases $i$ except for the end-exhalation phase; the end-exhalation phase image information was inserted into the artifact area defined by $\tilde{\mathbf{M}}$. Artifacts were only simulated within the lung area (i.e., for axial slices within the lung $z$-range defined by the lung bounding boxes, see \cref{Sec:data}), which was the focus of the present work.

\subsection{DL-based artifact detection}
\label{Sec:AF_detection}
Identification of 4D CT artifacts was performed by dedicated artifact detection networks trained to classify an axial CT slice as artifact-free or affected by an INT artifact (network 1) or by a DS artifact (network 2); characteristics and the general training process were similar for networks 1 and 2. 

For both applications, fully convolutional networks were trained (loss function: binary cross entropy; optimizer: Adam with learning rate of $1\times10^{-3}$; early stopping strategy; image intensities normalized to the range $[0,1]$ prior to training). The goal was to efficiently learn two mappings 
$\psi^{(\cdot)}:\mathbf{I}\mapsto\psi^{(\cdot)}(\mathbf{I})\in\mathbb{R}^{n_z}$, $(\cdot)\in\{\text{INT, DS}\}$,  
that output for an input phase CT 
$\mathbf{I}$
a vector $\psi^{\text{INT}}(\mathbf{I})\in\mathbb{R}^{n_z}$ and $\psi^{\text{DS}}(\mathbf{I})\in\mathbb{R}^{n_z}$, respectively. The output vectors indicate the probability of the axial slice at position $z\in[1,n_z]$ to be artifact-affected. 
In line with \cite{madesta:2018}, to reduce GPU memory requirements as well as network training and inference time, networks where trained on coronal CT slices. During inference, to obtain a robust estimation for a 3D CT image $\mathbf{I}$, predictions $\psi_x^{(\cdot)}(\mathbf{I}_y)\in\mathbb{R}^{n_z}$ were computed for all coronal slices $\mathbf{I}_y\in\mathbb{Z}^{n_x\times n_z}$, $y\in\{1,n_y\}$, of $\mathbf{I}$ and its lung bounding box, respectively. The $z$-th entry of the sought $\psi^{(\cdot)}(\mathbf{I})$ is the maximum of the $z$-th elements of the slice-wise predictions $\psi_x$, that is, if the axial slice at position $z$ is predicted to be  artifact-affected based on at least one coronal slice, it is considered artifact-affected.

The \textcolor{black}{general} architecture of the network\textcolor{black}{s} to realize the mapping\textcolor{black}{s} $\psi_x$ is shown in \cref{fig:fig_3}, \textcolor{black}{Block 0}. The selection of nine blocks was due to the standard in-plane CT slice size of $n_x \times n_y= 512\times512$ pixels. Thus, nine pooling steps with the pooling being applied only along the $y$-axis lead to the desired output vector of size $1\times n_z\equiv n_z$.
For the vectors $\psi^{\text{INT}}(\mathbf{I}_{\mathrm{A}})$ and $\psi^{\text{DS}}(\mathbf{I}_{\mathrm{A}})$ obtained for an artifact-affected CT image $\mathbf{I}_{\mathrm{A}}$ by \textcolor{black}{application of the final artifact type-specific networks 1 and 2}, as a final step, binary, artifact \textcolor{black}{type-specific} masks \textcolor{black}{$\mathbf{M}_{\mathrm{A}}^{\text{INT}}$ and $\mathbf{M}_{\mathrm{A}}^{\text{DS}}$} were computed (1's for all voxels of the axial slices indicated to be artifact-affected; 0's for all other slices).

\subsection{DL-based conditional inpainting}
\label{Sec:AF_inpainting}
\begin{figure*}
    \centering
    \includegraphics[]{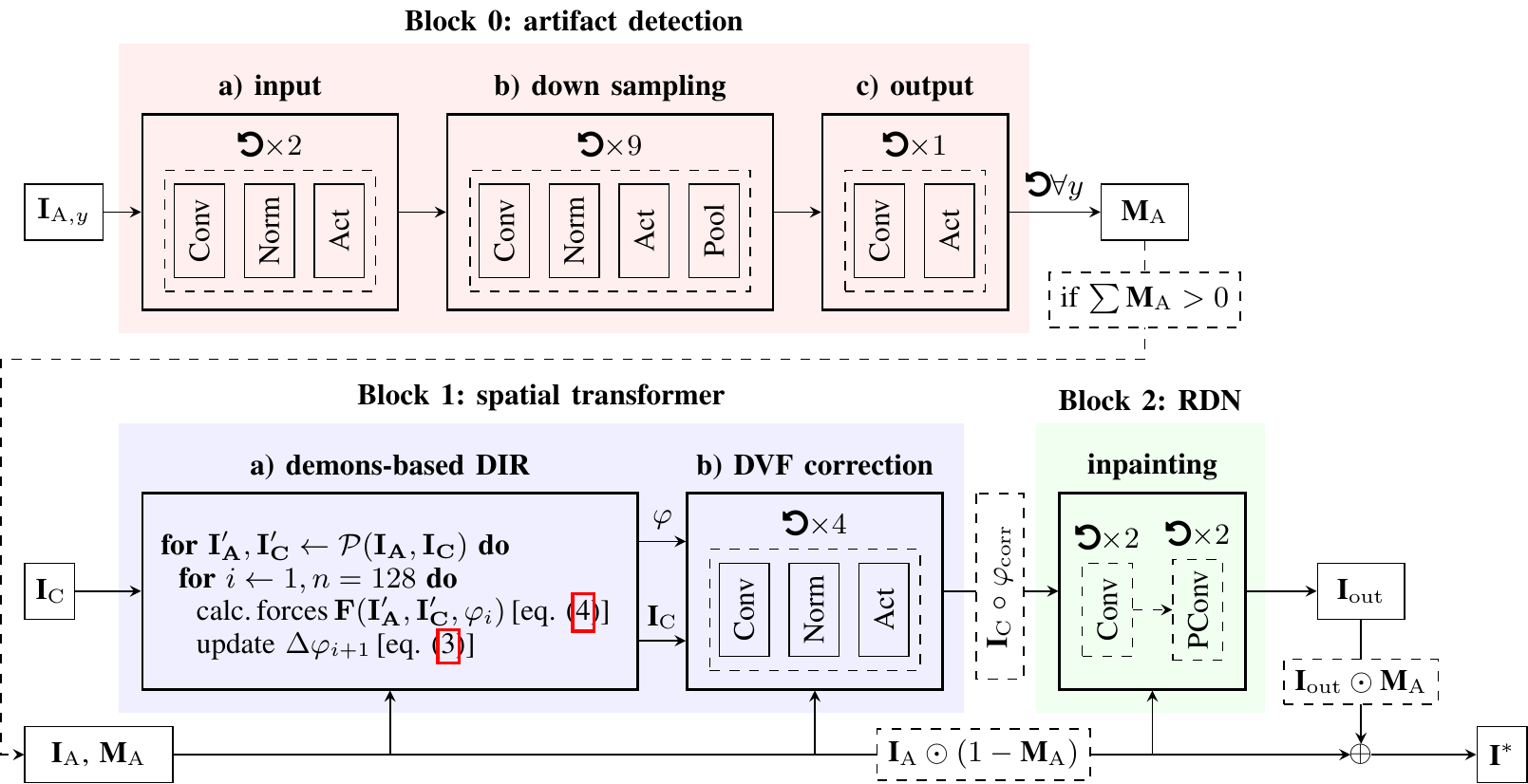}
    \caption{Flowchart of our proposed artifact detection and inpainting approach. \textcolor{black}{Note that the information flow and the network structure is similar for INT and DS artifacts. However, different conditional images $\mathbf{I}_\mathrm{C}$ are used and artifact type-specific networks are trained. For both artifact types,} prior to inpainting, an artifact \textcolor{black}{type-specific} mask  $\mathbf{M}_\mathrm{A}$ is predicted for each 3D phase CT by application of an artifact detection network (normalization: batch norm; activation [block 0(a) and (b)]: ReLU; pooling: max pool, activation [block 0(c)]: sigmoid). The detection is performed slice-wise for all coronal CT slices $\mathbf{I}_{A,y}$ overlapping with the lung bounding box of the 3D CT image. The results are combined as detailed in the main text; the output is a 3-dimensional binary mask $\mathbf{M}_\mathrm{A}$ (1: artifact, 0: else). Subsequently, the conditional image $\mathbf{I}_\mathrm{C}$, artifact-affected image $\mathbf{I}_\mathrm{A}$ and the predicted $\mathbf{M}_\mathrm{A}$ are fed into block 1, the spatial transformer block. In sub-block (a), a vector field $\varphi$ that aims at maximizing the similarity of $\mathbf{I}_\mathrm{C}$ and $\mathbf{I}_\mathrm{A}\odot(\textbf{1}-\mathbf{M}_\mathrm{A})$ is estimated by demons-based DIR ($\mathcal{P}$: image pyramid; for parameter details see main text). An end-to-end trainable vector field correction network (activation: Mish, normalization: instance norm), as illustrated in sub-block (b), improves the resulting transformation by predicting a vector field correction $\Delta\varphi_\mathrm{corr}$, resulting in a final corrected transformation $\varphi_\mathrm{corr}$. Finally, artifact inpainting is performed in block 2 using a residual dense net (RDN) with inputs being the transformed conditional image ($\mathbf{I}_\mathrm{C}\circ\varphi$), $\mathbf{I}_\mathrm{A}$ and corresponding $\mathbf{M}_\mathrm{A}$. Here, the first sub-block (feature extend block, followed by two residual dense blocks with standard convolutions and a feature fuse block) extracts feature maps based on the masked artifact-affected image and the warped conditional image, which are employed in the second sub-block (similar structure, but using partial convolutions) to inpaint the artifact region defined by $\mathbf{M}_\mathrm{A}$.
    }
    \label{fig:fig_3}
\end{figure*}

The proposed general DL-based inpainting concept is similar for the two artifact types, but separate networks were trained for each type. Since the network\textcolor{black}{s} can be applied to 3D phase CTs of arbitrary breathing phases, the breathing phase subscript introduced in \cref{Sec:AF_simulation} is omitted in the following. The general network structure is sketched in \cref{fig:fig_3}\textcolor{black}{, blocks 1 and 2}. Inputs are the artifact image $\mathbf{I}_\mathrm{A}$, the corresponding artifact mask $\mathbf{M}_\mathrm{A}$, and a so-called conditional image $\mathbf{I}_\mathrm{C}$. The idea is that $\mathbf{I}_\mathrm{C}$ contributes correct anatomical information for the artifact-affected region, and thereby guides the inpainting process.
Due to the different causes and appearances of INT and DS artifacts in 4D CT data, different conditional images were selected for the two types. 

\emph{INT artifacts} usually affect specific slices in all phases of a 4D CT; thus, an individual phase image is not an appropriate conditional image. Therefore, the temporal average image associated with the 4D CT image (see \cref{Sec:data}) was used as conditional image. The temporal average CT is reconstructed without consideration of breathing phase information. It therefore contains no INT artifacts, but provides a motion-blurred representation of the anatomy.

For \emph{DS artifacts}, mainly phase images close(r) to end-inspiration are affected; the end-exhalation state is usually less affected by breathing variability and corresponding phase images are less prone to DS artifacts \cite{li2014,castillo2014}. We therefore selected the end-exhalation phase as the conditional image.

As illustrated in \cref{fig:fig_1}, 2D inpainting (that is, slice-wise inpainting) fails for larger motion amplitudes; thus, a 3D inpainting approach was developed. Yet, directly working on the entire 3D CT volumes is still challenging with standard hardware and GPU memory restrictions -- and also not necessary. We are only interested in adapting (ideally correcting) artifact-affected image areas, while leaving the other areas unchanged. Therefore, we implemented a 3D patch-based approach that focuses on the artifact areas indicated by $\mathbf{M}_\mathrm{A}$. In detail, for network training, 3D patches of size $p_x\times p_y\times p_z = 32\times64\times96$ voxels were defined and the central axial patch slices aligned with the central $z$-slices of the artifact area to be inpainted. Patch-boundary artifacts are suppressed by large patch dimensions and only using a patch center crop for reconstruction of the inpainted image. 
During network training, for a specific 3D phase image $\mathbf{I}_\mathrm{A}$ and the  conditional image $\mathbf{I}_\mathrm{C}$, patches were sampled at the same $z$-position determined by the artifact mask $\mathbf{M}_\mathrm{A}$; the $(x,y)$-position was randomly sampled within the lung bounding box. For a common image spacing of approximately 1\,mm $\times$ 1\,mm $\times$ 2\,mm, the selected patch size should provide sufficient context information even for larger motion-related shifts of anatomical structure positions between $\mathbf{I}_\mathrm{A}$ and $\mathbf{I}_\mathrm{C}$ but still warrants data efficient training even with limited GPU memory. During network inference after training, the patch size was increased to $96^3$ voxels to decrease overhead due to the applied sliding window inference by strided image crops.

In each case, the corresponding patches of $\mathbf{I}_\mathrm{A}$, $\mathbf{M}_\mathrm{A}$, and $\mathbf{I}_\mathrm{C}$ are input into the \textcolor{black}{artifact type-specific} inpainting network detailed in \cref{fig:fig_3}, with the individual blocks being explained below. 

\subsubsection{Spatial transformer block}
Even though the patches are sufficiently dimensioned to warrant that the desired anatomical information to be inpainted into the artifact area is in principle present in the conditional image patch, we hypothesized that anatomically correct alignment of the $\mathbf{I}_\mathrm{C}$ and $\mathbf{I}_\mathrm{A}$ patches before inpainting would allow improved guidance during inpainting.

To align the $\mathbf{I}_\mathrm{C}$ and $\mathbf{I}_\mathrm{A}$ patches, we propose a variant of the spatial transformer block \cite{jaderberg2015} that is based on the demons registration, which is known to be well-suited for motion estimation in 4D CT data \cite{werner:2014}. We  consider the $\mathbf{I}_\mathrm{A}$ patch as a deformably transformed and perturbed/artifact-affected version of the conditional image patch $\mathbf{I}_\mathrm{C}$, i.e.,
\begin{equation}
\mathbf{I}_\mathrm{A} = \mathbf{I}_\mathrm{C} \circ \varphi + \Delta\mathbf{I},
\end{equation}
where $\varphi = \mathrm{id} + \Delta\varphi$ and $\Delta\mathbf{I}$ denote the deformable transformation and a residual perturbation. To compute $\varphi$, we applied active demon forces, leading to the iterative update step
\begin{equation}
\label{eq:reg_update}
\Delta\varphi_{i+1} = \mathcal{R}\left(\Delta\varphi_i - \tau \mathbf{F}\right)
\end{equation}
with $\tau\in\mathbb{R}^+$ as time step (here: $\tau=2.25$) and voxel-wise force computation
\begin{equation}
\label{eq:demon_forces}
\mathbf{F}(\mathbf{I}_\mathrm{A}, \mathbf{I}_\mathrm{C}, \varphi_i) = \frac{\mathbf{I}_\mathrm{A}-\mathbf{I}_\mathrm{C}\circ\varphi_i}
{\|\nabla\mathbf{I}_\mathrm{C}\circ\varphi_i\|^2+\alpha(\mathbf{I}_\mathrm{A}-\mathbf{I}_\mathrm{C}\circ\varphi_i)^2}
\nabla\mathbf{I}_\mathrm{C}\circ\varphi_i
\end{equation}
with $\alpha\in\mathbb{R}^+$ as the inverse of the mean squared image spacing and $\mathcal{R}$ as Gaussian regularization, in our case with  $\sigma=(1.25, 1.25, 1.25)$ voxels. \textcolor{black}{The influence of the hyperparameters $\tau$ and $\sigma$ on the registration result are discussed in detail in \cite{werner:2014}.} 

$\Delta\varphi$ was intialized with a zero displacement field and forces were only computed for non-artifact voxels, i.e., force computation was masked by $(\mathbf{1}-\mathbf{M}_\mathrm{A})$. To account for larger motion amplitudes, the registration scheme was successively conducted for a two-level image pyramid $\mathcal{P}(\mathbf{I}_\mathrm{A}, \mathbf{I}_\mathrm{C})$ with image scales of 0.5 and 1.0. The transformation estimated on the coarser level acted as initialization for the finer one. 
On both levels, DIR was stopped once the relative change of the mean squared error between $\mathbf{I}_\mathrm{A}\odot(\textbf{1}-\mathbf{M}_\mathrm{A})$ and $\mathbf{I}_\mathrm{C}\circ\varphi_i$ was below $10^{-2}$ for 25 steps or when the maximum number of iterations, here 128, was reached.
Thus, $\varphi$ was computed to match $\mathbf{I}_\mathrm{C}$ and $\mathbf{I}_\mathrm{A}$, while only implicitly taking into account $\Delta\mathbf{I}$ by force computation masking. 

It is, however, known that image artifacts influence deformable image registration (DIR) and lead to locally disturbed displacement fields \cite{mogadas:2018,sothmann:2018}. We therefore combined the registration step with an end-to-end trainable vector field correction block, consisting of four convolution blocks and receiving a 5-channel input, consisting of the three channels of the computed vector field $\Delta \varphi$, the masked artifact image $\mathbf{I}_\mathrm{A}\odot(\textbf{1}-\mathbf{M}_\mathrm{A})$ and the warped conditional image $\mathbf{I}_\mathrm{C}\circ\varphi$. The output, a vector field correction $\Delta\varphi_\mathrm{corr}$, is then added to the initially computed vector field, followed by a final Gaussian regularization iteration, resulting in a corrected deformable transformation $\varphi_\mathrm{corr} = \mathrm{id} + \mathcal{R}(\Delta\varphi + \Delta\varphi_\mathrm{corr})$.

\subsubsection{Conditional inpainting using partial convolutions}
\label{sec:conditional_inpainting}
The inpainting concept applied in this study is motivated by Liu \textit{et al.} \cite{liu:2018}, who introduced the concept of partial convolutions (PConv) for inpainting of irregular holes in natural 2D images. The key idea of the authors was to separate missing (belonging to the holes) and valid pixels (other areas of the image) during convolution operations. The convolutions are therefore \textit{partially} performed on the input, constrained by a binary mask to highlight valid image pixels, and are followed by a mask update step to expand the mask area. Repeating this process results in missing pixel information of the input image being gradually filled in \cite{liu:2018}.

In our case, we assume that the image regions affected by artifacts provide invalid information and use partial convolutions to exclude contributions of these regions during inpainting. The remaining question was, however, how to provide guidance by $\mathbf{I}_\mathrm{C}$ when using partial convolutions. After extensive experiments, we chose the two-step approach sketched in \cref{fig:fig_3}. First, the warped conditional image $\mathbf{I}_\mathrm{C}\circ\varphi$ and masked artifact image $(\mathbf{1}-\mathbf{M}_\mathrm{A})\odot\mathbf{I}_\mathrm{A}$ were input into a residual dense network (RDN) with two residual dense blocks consisting of four standard, i.e., non-partial, convolution layers each. The RDN output was a 10-channel feature map, with each single channel feature map being of same size as the image patches. These were combined with the masked artifact image by concatenation, and the resulting 11-channel data was input into a second RDN, which was similar in structure but comprised only partial convolution layers.
At this, $(\mathbf{1}-\mathbf{M}_\mathrm{A})$ served as the mask to indicate valid input voxels for all channels. As the two blocks were trained end-to-end, the intention was to force the first RDN block to extract sufficient information from $\mathbf{I}_\mathrm{C}\circ\varphi$ and present it at the correct locations in the feature maps to successfully guide the partial inpainting process of the second RDN.

The output of the second RDN, $\mathbf{I}_\mathrm{out}$, was finally combined with the original input patch of $\mathbf{I}_\mathrm{A}$ to obtain the sought inpainted image patch by
\begin{equation}
    \label{eq:inpainting}
    \mathbf{I}^* = (\mathbf{1}-\mathbf{M}_\mathrm{A})\odot\mathbf{I}_\mathrm{A}+\mathbf{M}_\mathrm{A}\odot\mathbf{I}_\mathrm{out}
\end{equation}
to warrant that only the artifact areas were changed. 

In the original partial convolutions paper, Liu \textit{et al.} \cite{liu:2018} combined several loss terms (L1, perceptual, style and total variation loss) to achieve state-of-the-art inpainting performance for natural image inpainting, where perceptual, style and total variation loss terms generally aim at generating a smooth contextual image impression. For inpainting of artifacts in the medical image domain, however, there exists only one constraint: anatomical correctness. We therefore considered only pixel-wise reconstruction accuracy, i.e., the L1 loss, as loss function for training.  

\subsection{Experiments and evaluation strategy}
\label{sec:experiments}
The evaluation was split into three experimental parts. For the first two experiments, that is, a quantitative evaluation of the artifact detection and inpainting performance, the following data were employed: 12 of the 21 artifact-free \textcolor{black}{in-house} 4D CT data sets (cf. \cref{Sec:data}) were randomly picked and served as validation data set, avoiding interference of original and simulated artifacts during evaluation. The remaining 9 artifact-free and the 27 slightly affected data sets were merged and used as training data. Thus, the experiments were conducted with a 75\%/25\% train/validation split (36/12 4D CT data sets). 
\textcolor{black}{In addition, the 10 DIR-Lab cases and the slightly artifact-affected/artifact-free cases of the 4D-Lung cohort (8 of 18 4D CT data sets) served as independent external test set. The third experiment aimed to demonstrate applicability to clinical artifact-affected data, making use of the in-house 4D CT data and the remaining 10 cases of the 4D-Lung data with severe artifacts.}
\begin{figure*}
    \centering
    \includegraphics[]{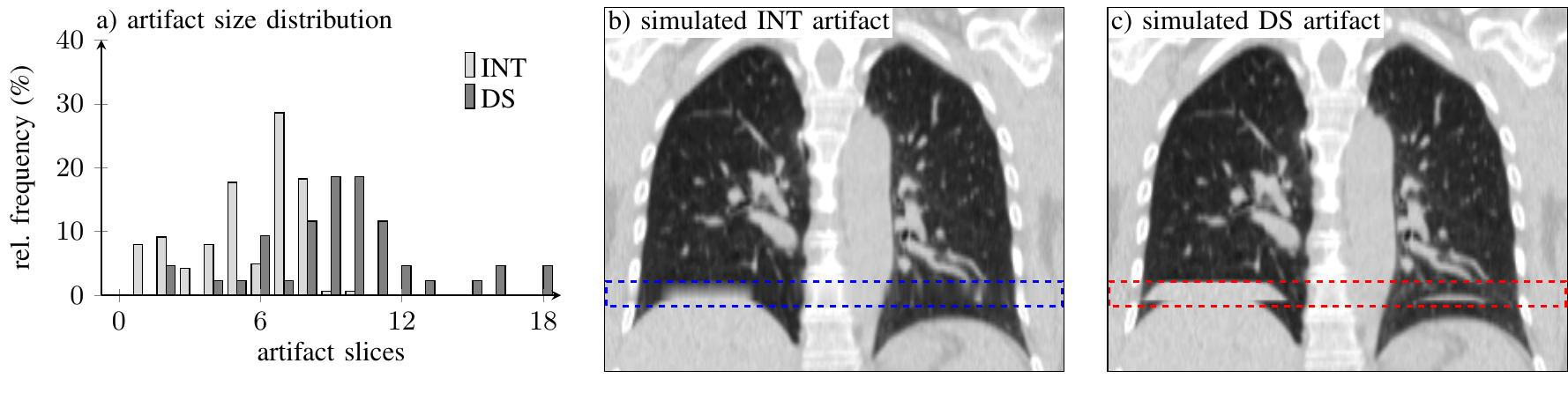}
    \caption{a) INT and DS artifact size distributions for a subset of the in-house data set. b) Illustration of simulated INT \textcolor{black}{(highlighted by blue box)} and c) DS \textcolor{black}{(red box)} artifacts.}
    \label{fig:fig_2}
\end{figure*}

\subsubsection{Experiment I: Artifact detection accuracy.} The \textcolor{black}{artifact type-specific} model\textcolor{black}{s} presented in \cref{Sec:AF_detection} \textcolor{black}{were} trained and tested using the aforementioned data split.
Based on an initial evaluation of the in-house data set, we randomly inserted up to three (95th percentile of the number of artifacts per 4D CT phase image) DS or INT artifacts into a single 4D CT phase image according to \cref{Sec:AF_simulation}. Simulated artifacts are illustrated in \cref{fig:fig_2}\,b and c. 
INT artifacts were simulated for axial slices along the entire $z$-range of the lung bounding box; DS artifacts were only generated in the lower third of the bounding box as they usually occur in areas of more pronounced structure motion, i.e., in lower parts of the lung.  
The artifact size was sampled from a uniform distribution in the range of $[2, 18]$ slices (see \cref{fig:fig_2}\,a). All available 4D CT phase images were used for the artifact simulation.
The model performance was evaluated using standard accuracy (patient-wise evaluation, weighted by number of artifact slices) and AUC-ROC measures on all 4D CT phase images in the validation set.
\subsubsection{Experiment II: Artifact inpainting performance (simulated artifacts)} The artifact inpainting model (cf. \cref{Sec:AF_inpainting}) was trained and validated using the aforementioned image data set splits. 
For experiment II, a single artifact was inserted into the 4D CT data on the phase image level. Similar to experiment I, the artifact size and location was sampled from the stated uniform distribution and every available 4D CT phase image was used for simulating artifacts during training. 
For each 4D CT phase, 100 training patches were extracted from each simulated artifact image $\mathbf{I}_\mathrm{A}$ and conditional image $\mathbf{I}_\mathrm{C}$, respectively. The (nearly) artifact-free ground truth patches were used to calculate the L1 loss during training.

\paragraph{Ablation study} First, to investigate the influence of the individual building blocks of our inpainting approach (cf. \cref{fig:fig_3}), an ablation study with the following configurations was conducted:

\textcolor{black}{\noindent\textsc{Inpaint}\,[$\mathbf{I}_\mathrm{C}\circ\varphi_\text{corr}$] (i.e., \textsc{Full model}):} Both the spatial transformer block and the RDN block are enabled to the full extent, i.e., demons-based DIR, DVF correction as well as the PConv-based RDN are used. The latter receives the conditional image $\mathbf{I}_\mathrm{C}\circ\varphi_\mathrm{corr}$ warped with the corrected deformable transformation $\varphi_\mathrm{corr}$, the masked artifact image $(\mathbf{1}-\mathbf{M}_\mathrm{A})\odot\mathbf{I}_\mathrm{A}$ as well as the artifact mask $\mathbf{M}_\mathrm{A}$ mandatory for the partial convolutions. 

\textcolor{black}{\noindent\textsc{Inpaint}\,[$\mathbf{I}_\mathrm{C}\circ\varphi$]:} Same as the \textsc{Full model} but with the DVF correction switched off (uncorrected DIR for computing $\varphi$ remains enabled).

\textcolor{black}{\noindent\textsc{Inpaint}\,[$\mathbf{I}_\mathrm{C}$]:} Same as the \textsc{Full model} but with the DIR completely switched off, i.e., no warping is applied to the conditional image $\mathbf{I}_\mathrm{C}$ ($\varphi = \mathrm{id}$).

\textcolor{black}{\noindent\textsc{Fill}\,[$\mathbf{I}_\mathrm{C}\circ\varphi_\text{corr}$], \textsc{Fill}\,[$\mathbf{I}_\mathrm{C}\circ\varphi$] and \textsc{Fill}\,[$\mathbf{I}_\mathrm{C}$]: Same as the respective \textsc{Inpaint} approach but with the inpainting block (\cref{fig:fig_3}, green) switched off, i.e., the network output is the given conditional image warped with the corrected DVF ($\mathbf{I}_\mathrm{C}\circ\varphi_\mathrm{corr}$), the uncorrected DVF ($\mathbf{I}_\mathrm{C}\circ\varphi$) and without warping ($\varphi = \mathrm{id}$), respectively.}

\textcolor{black}{\noindent\textsc{Inpaint}\,[$\mathbf{I}_\mathrm{A}$]:} To investigate the effect of $\mathbf{I}_\mathrm{C}$, we completely disregard the conditional image $\mathbf{I}_\mathrm{C}$ and perform standard inpainting with just the masked artifact image $(\mathbf{1}-\mathbf{M}_\mathrm{A})\odot\mathbf{I}_\mathrm{A}$ and the artifact mask $\mathbf{M}_\mathrm{A}$ directly fed into the RDN inpainting block.

In addition, the \textsc{Full model} configuration was trained with partial convolutions and standard convolutions to evaluate the impact of the proposed use of PConv on the model performance. 
\textcolor{black}{For each configuration, a separate model was trained to allow for a fair and unbiased comparison.} 
\textcolor{black}{
If a model did not improve in a patience window of 10 epochs, we assumed convergence and the best performing model so far (based on the loss function for the validation data) was selected as the final model for the respective configuration.} 
\textcolor{black}{The performance of the different  configurations was evaluated for the in-house validation data set.}

\paragraph{\textcolor{black}{Generalizability and comparison to state-of-the-art}} 
\textcolor{black}{To analyze the generalizability of the trained models to external 4D CT data, the trained \textsc{Full model} was applied to the external test sets with simulated artifacts. 
To be able to assess the model performance, we compared the performance to the \textsc{Inpaint}\,[$\mathbf{I}_\mathrm{A}$] configuration (i.e., plain inpainting) as a baseline model and to two recent approaches in the given context. The first method was a recently published state-of-the-art registration-based approach to the reduction of 4D CT image artifacts, namely GDR \cite{shao:2021}. GDR was applied by directly using the respective openly available code base along with the corresponding parameters.} 
\textcolor{black}{To compare the performance of the proposed DL-based method to a GAN-based inpainting approach, we adopted the model developed by Jin \textit{et al.} \cite{jin:2021}. The choice was motivated by the 3D nature of their implementation, the good performance of the model when applied to thoracic CT data for the original task (generation of synthetic lesion), and the publicly available code. After the first few training epochs with only L1 loss to stabilize training, we continued model training with the parameters recommended by the authors. Training data as well as the definition of model convergence were identical to the training of our models.}

\textcolor{black}{In addition, we evaluated the performance of our \textsc{Full model} when trained on only a subset of the training data to investigate the impact of the size of the training data set on model performance. For this substudy, we randomly selected 25\%, 50\% and 75\% of the full training data set for model training and, based on the external test sets, compared the performance of the corresponding models \textsc{Full model$_{25\%}$}, \textsc{Full model$_{50\%}$}, and \textsc{Full model$_{75\%}$} to the \textsc{Full model} of experiment I.}   

For the in-house validation set as well as the external 4D CT data sets, inpainting performance was quantitatively evaluated by the root mean squared error (RMSE) and the normalized cross correlation (NCC) between the original (almost) artifact-free phase CT $\mathbf{I}$ of the 4D CT data and the corresponding inpainting result $\mathbf{I}^*$ after processing the artifact-affected version $\mathbf{I}_\mathrm{A}$ of $\mathbf{I}$ (cf. \eqref{eq:inpainting}).  
The computation of RMSE and NCC was restricted to an evaluation mask $\mathbf{M}_\mathrm{E}$, defined as the intersection of the total lung mask $\mathbf{M}_\mathrm{L}$ and the artifact mask $\mathbf{M}_\mathrm{A}$; the total lung mask was the union of the 10 lung masks of the considered 10-phase 4D CT.

For all models and configurations, the performance was analyzed separately for the different artifact sizes to allow for a detailed analysis of the influence of the artifact size on the inpainting performance. 	
Since the most pronounced artifacts are those where end-exhale slices were inserted into an end-inhale phase image (i.e., $i=1$ and $j=5$, cf. \eqref{eq:AF_simulation}), the evaluation focused on exactly this artifact simulation setting, i.e., the worst-case scenario, during evaluation. To cover a wide variety of artifact location/size combinations, the whole testing was performed 10 times \textcolor{black}{for the DL-based models. For GDR, the excessive computation time forced us to reduce the number of testing runs to a single run per patient and artifact size.}
\subsubsection{Experiment III: End-to-end validation on clinical artifact-affected data} \textcolor{black}{The entire inpainting workflow (artifact detection models trained for experiment~I; \textsc{Full model} trained during experiment~II) was applied to the 18 in-house and the 10 4D-Lung 4D CT data sets with pronounced artifacts (see \cref{Sec:data}).} 
As no ground truth was available in this case, the inpainting performance was quantitatively evaluated by applying the artifact detection models to the original and the processed data and analyzing the resulting detection network outputs before and after inpainting.

% ---------- RESULTS SECTION:
\begin{figure*}
    \centering
    \includegraphics[]{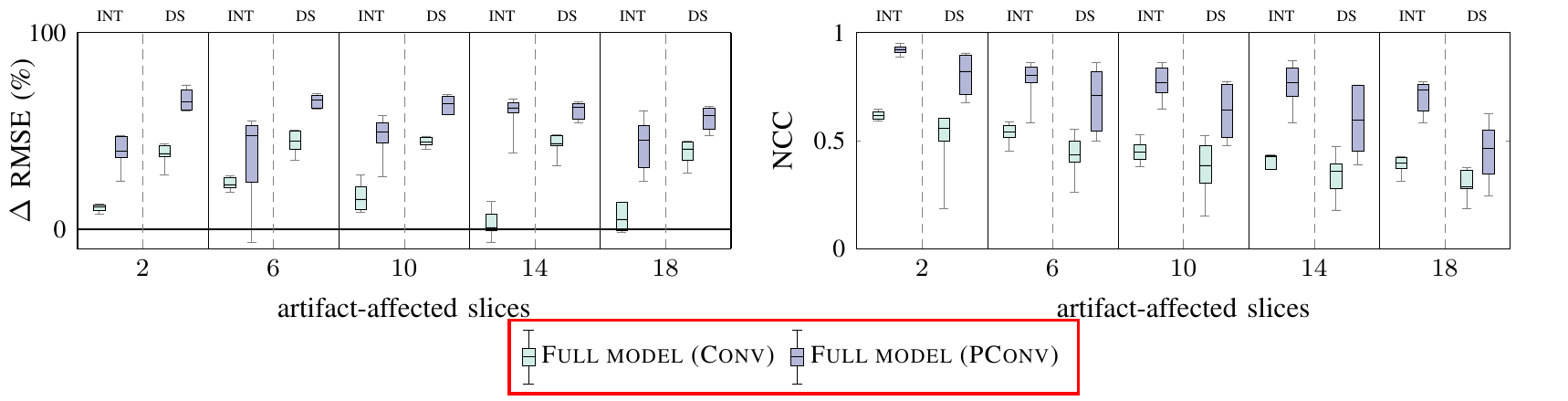}
    \caption{Inpainting performance of the \textsc{Full model} configurations with Conv/PConv layers for simulated artifacts and the in-house validation data, quantified by $\Delta$RMSE (left) and NCC (right) values.}
    \label{fig:results_conv_vs_pconv_ncc+rmse}
\end{figure*}

\section{Results}
\subsection{Artifact detection}
In experiment I, artifact detection networks for INT artifacts (network~1) and DS artifacts (network~2) were evaluated. For network~1, mean ROC-AUC of 0.99 and accuracy of 0.97 (sensitivity 0.97, specificity 0.97; optimal operating point by Youden's Index) indicate a high detection probability of INT artifacts. 

The evaluation of network~2 reveals similar numbers for ROC-AUC (0.97) and sensitivity (0.96) at the optimal operating point.
However, in comparison to network~1, the accuracy and specificity decreases (0.86 and 0.85, respectively), suggesting that DS artifacts are detectable, but some slices of data originally labeled as artifact-free are classified as DS artifact-affected. A visualization of detected artifacts is given for the end-to-end evaluation in \cref{fig:inhouse_end_to_end} (left column).

\subsection{Artifact inpainting}
The general effect and performance of the proposed inpainting approach was evaluated as described in experiment II, i.e., based on artifact-free data that serve as ground truth images, and corresponding images with simulated artifacts. In the following, inpainting performance is expressed by RMSE improvement ($\Delta$RMSE; i.e., RMSE after inpainting compared to RMSE before inpainting) and absolute NCC values.

The results \textcolor{black}{of the comparison of Conv and PConv layers for the given task are summarized} in \cref{fig:results_conv_vs_pconv_ncc+rmse} and indicate that employing PConv layers for inpainting is mandatory to achieve high similarity of the inpainted and the ground truth CT image data in particular for INT artifacts. 
\textcolor{black}{Using only Conv layers marginally improved the image quality compared to the INT artifact-affected input images (near zero median $\Delta$RMSE values for artifact sizes of 2, 14 and 18 slices, cf. left plot in \cref{fig:results_conv_vs_pconv_ncc+rmse}),
while the application of PConv layers led to 51\% image similarity improvement in terms of RMSE, averaged over all artifacts sizes. 
For DS artifacts, inpainting with PConv layers consistently outperformed Conv layers for all evaluated artifact sizes (Conv/PConv: average $\Delta$RMSE: 43\,\%/59\,\%, NCC: 0.41/0.65).}

To illustrate the effect of the different building blocks of the proposed inpainting pipeline, the inpainting performance of the configurations detailed in \cref{sec:experiments} 
\textcolor{black}{are summarized in \cref{tab:results_ablation_study_in-house}.} 
For both INT and DS artifacts, the \textsc{Full model} configuration achieved \textcolor{black}{consistently} the best inpainting result, \textcolor{black}{for all artifact sizes} and averaged over the full range of artifact sizes (INT/DS: average RMSE improvement  \textcolor{black}{52}\,\%/\textcolor{black}{59}\,\%, average NCC: \textcolor{black}{0.80}/\textcolor{black}{0.65}). 
\textcolor{black}{Moreover, comparing \textcolor{black}{\textsc{Inpaint}\,[$\mathbf{I}_\mathrm{C}$]} and \textcolor{black}{\textsc{Inpaint}\,[$\mathbf{I}_\mathrm{C}\circ\varphi$] (approaches d and e in \cref{tab:results_ablation_study_in-house})} for larger artifact sizes shows that enabling registration without correction has already a large positive impact on the overall performance compared to not performing a registration at all.}
\textcolor{black}{The configurations with disabled inpainting block, i.e., \textsc{Fill}\,[$\mathbf{I}_\mathrm{C}$], \textsc{Fill}\,[$\mathbf{I}_\mathrm{C}\circ\varphi$] and \textsc{Fill}\,[$\mathbf{I}_\mathrm{C}\circ\varphi_{\text{corr}}$], show the same tendency but inferior performance compared to their corresponding \textsc{Inpaint} method.}

\begin{table}
\centering
\scriptsize
\color{black}\caption{Ablation study: inpainting performance for our \textsc{Fill}\,[$\mathbf{I}_\mathrm{C}$], \textsc{Fill}\,[$\mathbf{I}_\mathrm{C}\circ\varphi$], \textsc{Fill}\,[$\mathbf{I}_\mathrm{C}\circ\varphi_{\text{corr}}$], \textsc{Inpaint}\,[$\mathbf{I}_\mathrm{C}$], \textsc{Inpaint}\,[$\mathbf{I}_\mathrm{C}\circ\varphi$] and the \textsc{Full model} configurations for simulated INT and DS artifacts and the in-house validation data (quantified by $\Delta$RMSE and NCC values; numbers in the format $a^b_c$, with $a$ denoting the median, $b$ the 75\% percentile and $c$ the 25\% percentile).}
\label{tab:results_ablation_study_in-house}
\color{black}\scalebox{0.9}{
\begin{tabular}{wl{0.25cm}wl{0.25cm}wl{0.25cm}*{5}{C{1.1cm}}}
\toprule
\multicolumn{3}{c}{ \# \textbf{slices:}} & \textbf{2} & \textbf{6} & \textbf{10} & \textbf{14} & \textbf{18} \\
\midrule
\multirow{18}{*}{\rotatebox[origin=c]{90}{\textbf{INT artifacts}}} &
\multirow{9}{*}{\rotatebox[origin=c]{90}{$\mathbf{\Delta}$\textbf{RMSE (\%)}}} 
& a & $-$61.0$_\texttt{-77.4}^\texttt{-44.1}$	&	$-$14.4$_\texttt{-37.8}^\texttt{~22.1}$	&	\phantom{$-$}15.6$_\texttt{~11.8}^\texttt{~31.6}$	&	\phantom{$-$}35.6$_\texttt{~14.5}^\texttt{~40.1}$	&	\phantom{$-$}29.5$_\texttt{~25.4}^\texttt{~35.5}$ \\ \addlinespace
&& b & \phantom{$-$}\phantom{0}2.2$_\texttt{-19.6}^\texttt{~15.9}$	&	\phantom{$-$}27.1$_\texttt{~14.1}^\texttt{~48.2}$	&	\phantom{$-$}40.7$_\texttt{~24.0}^\texttt{~52.6}$	&	\phantom{$-$}54.3$_\texttt{~35.5}^\texttt{~56.9}$	&	\phantom{$-$}51.1$_\texttt{~39.5}^\texttt{~57.6}$ \\ \addlinespace
&& c & \phantom{$-$}10.8$_\texttt{-~1.1}^\texttt{~27.2}$	&	\phantom{$-$}36.0$_\texttt{~23.5}^\texttt{~55.1}$	&	\phantom{$-$}43.8$_\texttt{~34.7}^\texttt{~58.3}$	&	\phantom{$-$}56.8$_\texttt{~38.2}^\texttt{~58.4}$	&	\phantom{$-$}53.2$_\texttt{~42.1}^\texttt{~63.2}$ \\ \addlinespace
&& d & \phantom{$-$}37.1$_\texttt{~24.1}^\texttt{~43.0}$	&	\phantom{$-$}28.7$_\texttt{~25.1}^\texttt{~42.7}$	&	\phantom{$-$}47.2$_\texttt{~17.1}^\texttt{~54.5}$	&	\phantom{$-$}47.3$_\texttt{~28.7}^\texttt{~55.9}$	&	\phantom{$-$}24.5$_\texttt{~~9.9}^\texttt{~35.6}$ \\ \addlinespace
&& e & \phantom{$-$}32.9$_\texttt{~28.7}^\texttt{~39.5}$	&	\phantom{$-$}43.6$_\texttt{~35.0}^\texttt{~53.9}$	&	\phantom{$-$}54.3$_\texttt{~48.2}^\texttt{~59.7}$	&	\phantom{$-$}56.2$_\texttt{~45.6}^\texttt{~60.9}$	&	\phantom{$-$}57.0$_\texttt{~49.9}^\texttt{~64.5}$ \\ \addlinespace
&& f & \bftab\phantom{$-$}39.6$_\texttt{~34.7}^\texttt{~43.8}$	&	\bftab\phantom{$-$}45.5$_\texttt{~34.0}^\texttt{~53.2}$	&	\bftab\phantom{$-$}55.4$_\texttt{~45.0}^\texttt{~62.4}$	&	\bftab\phantom{$-$}60.1$_\texttt{~44.5}^\texttt{~62.4}$	&	\bftab\phantom{$-$}57.7$_\texttt{~51.7}^\texttt{~63.5}$ \\
\cmidrule{2-8}
&\multirow{9}{*}{\rotatebox[origin=c]{90}{\textbf{NCC}}} 
& a & \phantom{$-$}0.60$_\texttt{~0.51}^\texttt{~0.67}$	&	\phantom{$-$}0.51$_\texttt{~0.39}^\texttt{~0.62}$	&	\phantom{$-$}0.49$_\texttt{~0.44}^\texttt{~0.62}$	&	\phantom{$-$}0.60$_\texttt{~0.49}^\texttt{~0.68}$	&	\phantom{$-$}0.49$_\texttt{~0.44}^\texttt{~0.55}$ \\ \addlinespace
&& b & \phantom{$-$}0.80$_\texttt{~0.74}^\texttt{~0.84}$	&	\phantom{$-$}0.73$_\texttt{~0.60}^\texttt{~0.78}$	&	\phantom{$-$}0.67$_\texttt{~0.55}^\texttt{~0.81}$	&	\phantom{$-$}0.73$_\texttt{~0.63}^\texttt{~0.80}$	&	\phantom{$-$}0.67$_\texttt{~0.51}^\texttt{~0.78}$ \\ \addlinespace
&& c & \phantom{$-$}0.84$_\texttt{~0.81}^\texttt{~0.88}$	&	\phantom{$-$}0.77$_\texttt{~0.66}^\texttt{~0.82}$	&	\phantom{$-$}0.74$_\texttt{~0.59}^\texttt{~0.84}$	&	\phantom{$-$}0.76$_\texttt{~0.68}^\texttt{~0.83}$	&	\phantom{$-$}0.71$_\texttt{~0.53}^\texttt{~0.80}$ \\ \addlinespace
&& d & \phantom{$-$}0.89$_\texttt{~0.87}^\texttt{~0.90}$	&	\phantom{$-$}0.71$_\texttt{~0.68}^\texttt{~0.76}$	&	\phantom{$-$}0.69$_\texttt{~0.56}^\texttt{~0.71}$	&	\phantom{$-$}0.67$_\texttt{~0.60}^\texttt{~0.75}$	&	\phantom{$-$}0.62$_\texttt{~0.59}^\texttt{~0.66}$ \\ \addlinespace
&& e & \phantom{$-$}0.91$_\texttt{~0.90}^\texttt{~0.93}$	&	\phantom{$-$}0.78$_\texttt{~0.74}^\texttt{~0.84}$	&	\phantom{$-$}0.76$_\texttt{~0.74}^\texttt{~0.83}$	&	\phantom{$-$}0.75$_\texttt{~0.71}^\texttt{~0.82}$	&	\phantom{$-$}0.73$_\texttt{~0.65}^\texttt{~0.76}$ \\ \addlinespace
&& f & \bftab\phantom{$-$}0.92$_\texttt{~0.91}^\texttt{~0.94}$	&	\bftab\phantom{$-$}0.80$_\texttt{~0.77}^\texttt{~0.84}$	&	\bftab\phantom{$-$}0.77$_\texttt{~0.72}^\texttt{~0.84}$	&	\bftab\phantom{$-$}0.77$_\texttt{~0.71}^\texttt{~0.84}$	&	\bftab\phantom{$-$}0.74$_\texttt{~0.64}^\texttt{~0.76}$ \\
\cmidrule[\heavyrulewidth]{1-8}
\multirow{18}{*}{\rotatebox[origin=c]{90}{\textbf{DS artifacts}}} &
\multirow{9}{*}{\rotatebox[origin=c]{90}{$\mathbf{\Delta}$\textbf{RMSE (\%)}}} 
& a & $-$82.9$_\texttt{-85.8}^\texttt{-81.8}$	&	$-$26.8$_\texttt{-28.1}^\texttt{-23.6}$	&	$-$13.8$_\texttt{-15.4}^\texttt{-12.4}$	&	$-$\phantom{0}9.3$_\texttt{-13.8}^\texttt{-~8.1}$	&	$-$\phantom{0}7.8$_\texttt{-11.3}^\texttt{-~6.1}$ \\ \addlinespace
&& b & \phantom{$-$}47.4$_\texttt{~40.1}^\texttt{~56.1}$	&	\phantom{$-$}49.4$_\texttt{~32.3}^\texttt{~57.3}$	&	\phantom{$-$}54.9$_\texttt{~45.9}^\texttt{~61.0}$	&	\phantom{$-$}46.9$_\texttt{~38.7}^\texttt{~53.7}$	&	\phantom{$-$}41.6$_\texttt{~29.0}^\texttt{~49.3}$ \\ \addlinespace
&& c & \phantom{$-$}43.2$_\texttt{~33.9}^\texttt{~47.2}$	&	\phantom{$-$}47.6$_\texttt{~37.5}^\texttt{~60.1}$	&	\phantom{$-$}56.5$_\texttt{~46.3}^\texttt{~62.3}$	&	\phantom{$-$}49.7$_\texttt{~40.6}^\texttt{~52.4}$	&	\phantom{$-$}43.0$_\texttt{~35.1}^\texttt{~51.6}$ \\ \addlinespace
&& d & \phantom{$-$}31.3$_\texttt{~14.0}^\texttt{~46.4}$	&	\phantom{$-$}18.8$_\texttt{~~0.5}^\texttt{~33.9}$	&	\phantom{$-$}20.5$_\texttt{~12.1}^\texttt{~50.7}$	&	\phantom{$-$}25.9$_\texttt{~~1.3}^\texttt{~43.8}$	&	\phantom{$-$}28.5$_\texttt{~~6.7}^\texttt{~39.1}$ \\ \addlinespace
&& e & \phantom{$-$}57.4$_\texttt{~52.0}^\texttt{~61.8}$	&	\phantom{$-$}61.6$_\texttt{~59.7}^\texttt{~64.0}$	&	\phantom{$-$}60.4$_\texttt{~57.8}^\texttt{~63.0}$	&	\phantom{$-$}53.3$_\texttt{~52.3}^\texttt{~61.6}$	&	\phantom{$-$}52.9$_\texttt{~48.3}^\texttt{~59.5}$ \\ \addlinespace
&& f & \bftab\phantom{$-$}59.9$_\texttt{~56.5}^\texttt{~64.2}$	&	\bftab\phantom{$-$}62.7$_\texttt{~61.7}^\texttt{~68.3}$	&	\bftab\phantom{$-$}62.2$_\texttt{~59.2}^\texttt{~68.6}$	&	\bftab\phantom{$-$}57.3$_\texttt{~53.0}^\texttt{~66.5}$	&	\bftab\phantom{$-$}54.9$_\texttt{~49.4}^\texttt{~62.7}$ \\
\cmidrule{2-8}
&\multirow{9}{*}{\rotatebox[origin=c]{90}{\textbf{NCC}}} 
& a & \phantom{$-$}0.05$_\texttt{~0.01}^\texttt{~0.18}$	&	\phantom{$-$}0.05$_\texttt{~0.03}^\texttt{~0.14}$	&	\phantom{$-$}0.06$_\texttt{~0.03}^\texttt{~0.16}$	&	\phantom{$-$}0.05$_\texttt{~0.03}^\texttt{~0.12}$	&	\phantom{$-$}0.04$_\texttt{~0.01}^\texttt{~0.13}$ \\ \addlinespace
&& b & \phantom{$-$}0.71$_\texttt{~0.64}^\texttt{~0.85}$	&	\phantom{$-$}0.64$_\texttt{~0.48}^\texttt{~0.72}$	&	\phantom{$-$}0.58$_\texttt{~0.43}^\texttt{~0.71}$	&	\phantom{$-$}0.49$_\texttt{~0.33}^\texttt{~0.65}$	&	\phantom{$-$}0.41$_\texttt{~0.21}^\texttt{~0.50}$ \\ \addlinespace
&& c & \phantom{$-$}0.67$_\texttt{~0.58}^\texttt{~0.79}$	&	\phantom{$-$}0.63$_\texttt{~0.47}^\texttt{~0.74}$	&	\phantom{$-$}0.58$_\texttt{~0.44}^\texttt{~0.72}$	&	\phantom{$-$}0.50$_\texttt{~0.36}^\texttt{~0.67}$	&	\phantom{$-$}0.42$_\texttt{~0.23}^\texttt{~0.55}$ \\ \addlinespace
&& d & \phantom{$-$}0.52$_\texttt{~0.33}^\texttt{~0.64}$	&	\phantom{$-$}0.27$_\texttt{~0.21}^\texttt{~0.41}$	&	\phantom{$-$}0.19$_\texttt{~0.15}^\texttt{~0.33}$	&	\phantom{$-$}0.16$_\texttt{~0.13}^\texttt{~0.31}$	&	\phantom{$-$}0.15$_\texttt{~0.08}^\texttt{~0.24}$ \\ \addlinespace
&& e & \phantom{$-$}0.79$_\texttt{~0.71}^\texttt{~0.89}$	&	\phantom{$-$}0.62$_\texttt{~0.48}^\texttt{~0.80}$	&	\phantom{$-$}0.55$_\texttt{~0.46}^\texttt{~0.68}$	&	\phantom{$-$}0.54$_\texttt{~0.37}^\texttt{~0.72}$	&	\phantom{$-$}0.42$_\texttt{~0.28}^\texttt{~0.57}$ \\ \addlinespace
&& f & \bftab\phantom{$-$}0.82$_\texttt{~0.72}^\texttt{~0.90}$	&	\bftab\phantom{$-$}0.71$_\texttt{~0.55}^\texttt{~0.82}$	&	\bftab\phantom{$-$}0.64$_\texttt{~0.52}^\texttt{~0.76}$	&	\bftab\phantom{$-$}0.60$_\texttt{~0.45}^\texttt{~0.76}$	&	\bftab\phantom{$-$}0.47$_\texttt{~0.35}^\texttt{~0.55}$ \\
\bottomrule
\addlinespace
\multicolumn{8}{l}{a: \textsc{Fill}\,[$\mathbf{I}_\mathrm{C}$]\; b: \textsc{Fill}\,[$\mathbf{I}_\mathrm{C}\circ\varphi$]\; c: \textsc{Fill}\,[$\mathbf{I}_\mathrm{C}\circ\varphi_\text{corr}$]} \\
\multicolumn{8}{l}{d: \textsc{Inpaint}\,[$\mathbf{I}_\mathrm{C}$]\; e: \textsc{Inpaint}\,[$\mathbf{I}_\mathrm{C}\circ\varphi$]\; f: \textsc{Full model}}\\
\end{tabular}}
\end{table}
\begin{figure*}
    \centering
    \includegraphics[]{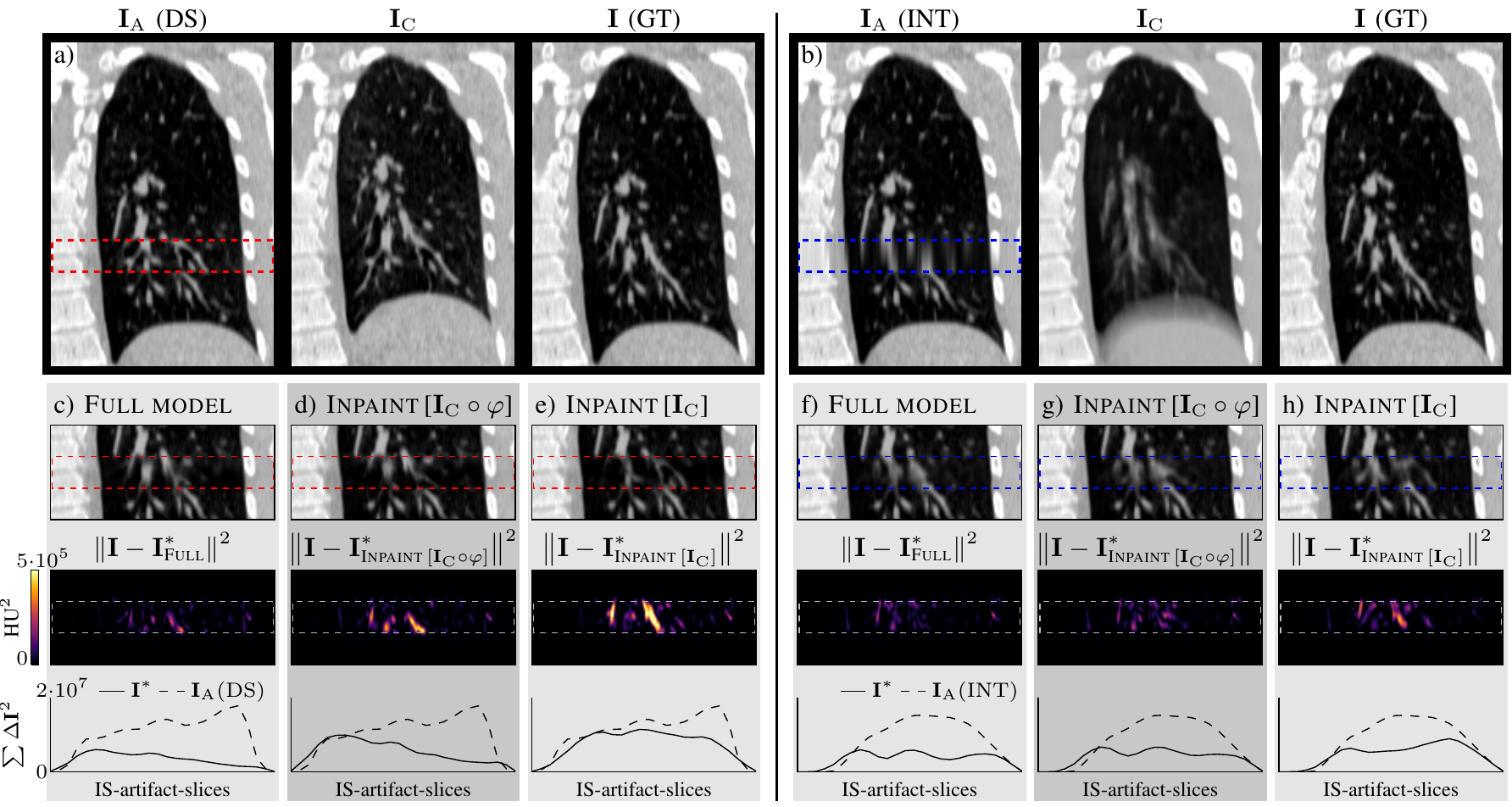}
   
    \caption{\textcolor{black}{DS/INT inpainting results for the configurations \textsc{Full model}, \textsc{Inpaint}\,[$\mathbf{I}_\mathrm{C}\circ\varphi$] and \textsc{Inpaint}\,[$\mathbf{I}_\mathrm{C}$] (DS: c--e/INT: f--h). The initial artifact images $\mathbf{I}_\mathrm{A}$ (artifact area marked in red/blue for DS/INT artifacts), conditional images $\mathbf{I}_\mathrm{C}$, and ground truth (GT), i.e., artifact-free images $\mathbf{I}$, are shown in blocks a and b for DS and INT artifacts, respectively. For each inpainting result, the squared intensity difference with respect to the GT image is shown (color-coded plots, areas outside the overlaid artifact masks are not modified, i.e., $\Delta \text{HU}^2 = 0$), and corresponding line profiles (summed differences in LR direction) are plotted in comparison to the artifact-affected image in the bottom row.}}
    \label{fig:ds_inpainting_results}
\end{figure*}

Figure \ref{fig:ds_inpainting_results} shows the inpainting results for the top 3 models (\textsc{Full model}, \textcolor{black}{\textsc{Inpaint}\,[$\mathbf{I}_\mathrm{C}$]}, \textcolor{black}{\textsc{Inpaint}\,[$\mathbf{I}_\mathrm{C}\circ\varphi$]}) for simulated artifacts of an exemplary data set of the artifact-free subgroup of our in-house validation data set. 
It can be seen that wrong lung structures are inserted if the spatial transformer block is completely disabled. Moreover, both \textcolor{black}{\textsc{Inpaint}\,[$\mathbf{I}_\mathrm{C}$]} and \textcolor{black}{\textsc{Inpaint}\,[$\mathbf{I}_\mathrm{C}\circ\varphi$]} resulted in a blurrier inpainting result than the \textsc{Full model} configuration. \textcolor{black}{However, compared to the ground truth image data, the \textsc{Full model} also leads to slight blurring of fine structures and structure borders.} Lastly, we would like to emphasize again that no image data is altered outside of the artifact area (red/blue rectangle in \cref{fig:ds_inpainting_results}) for any of our proposed inpainting methods.

\textcolor{black}{The \textsc{Full model} configuration was further evaluated on the external DIR-Lab and 4D-Lung data sets and compared to GDR and the 3D GAN inpainting approach (subsequently denoted as GAN). The corresponding results are shown in \cref{tab:results_ablation_study_dirlab} and \cref{tab:results_ablation_study_4d-lung}. Note that GDR is not able to inpaint INT artifacts (missing image information in all 4D CT phases). 
As an additional baseline, the \textsc{Inpaint}\,[$\mathbf{I}_\mathrm{A}$] configuration was applied to assess the performance of direct inpainting without guidance information by the conditional image $\mathbf{I}_\mathrm{C}$. 
The overall inpainting performance for the \textsc{Full model} was similar for the external test data sets as for the in-house data set (DIR-Lab: INT/DS, average RMSE improvement: 48\,\%/69\,\%, average NCC: 0.81/0.50; 4D-Lung: INT/DS, average RMSE improvement: 53\,\%/48\,\%, average NCC: 0.69/0.41). GDR requires at least one valid, i.e., artifact free phase image \cite{shao:2021}; this is not given for INT artifacts. For the DS artifacts, averaged over the two external data sets, GDR achieves RMSE improvements of 19\,\% and a mean NCC of near zero (0.08).}  
\textcolor{black}{The GAN-based approach performs particularly well for small DS artifacts, with the $\Delta$RMSE and NCC values for DS artifacts of size 2 in the DIR-Lab data set being better than the corresponding numbers of our \textsc{Full model}. With increasing DS artifact size, the GAN inpainting performance decreases and is (especially for the DIR-Lab data) with mean RMSE improvements and NCC values of 42\,\%/0.28 similar to our \textsc{Inpaint\,[$\mathbf{I}_\mathrm{A}$]} model (\textsc{Inpaint\,[$\mathbf{I}_\mathrm{A}$]}: 45\,\%/0.26). 
For INT artifacts, the GAN performs worse than \textsc{Inpaint\,[$\mathbf{I}_\mathrm{A}$]}, with a similar average NCC (GAN: 0.33; \textsc{Inpaint\,[$\mathbf{I}_\mathrm{A}$]}: 0.36) but a lower RMSE improvement (GAN: 1\,\%; \textsc{Inpaint\,[$\mathbf{I}_\mathrm{A}$]}: 20\,\%). 
The inpainting performance of GAN and \textsc{Inpaint\,[$\mathbf{I}_\mathrm{A}$]} demonstrates that the conditional image is in not leading to an increased inpainting performance for small INT and DS artifacts (2 slices), but the advantage of integrating additional patient-specific information becomes apparent with larger artifact sizes.}

\textcolor{black}{The analysis of the influence of the training set size on the performance of our approach revealed already high NCC but reduced $\Delta$RMSE values for the limited training sets. Relative to the corresponding values of the \textsc{Full model} that was trained on the entire training data set, the NCC values of the \textsc{Full model$_{25\%}$}, the \textsc{Full model$_{50\%}$}, and the \textsc{Full model$_{75\%}$} were reduced by 4.2\%, 5.8\%, and 3.2\% (averaged over INT and DS artifacts as well as the different artifact sizes). The $\Delta$RMSE values were reduced by 18.9\%, 15.2\%, and 6.1\%. Thus, the performance increases with a larger training set but seems to be relatively robust for the given training set size.}

\textcolor{black}{
The average inpainting wall clock time of the \textsc{Full model} was approximately 30\,s -- including CPU/GPU data transfer, deformable image registration, application of the inpainting model and reassembling of the resulting inpainted image. The inference speed was not dependent on the number of adjacent artifact-affected slices since the patch size in $z$-direction (96 voxel) covers the whole artifact in one step. However, if there are several artifacts, multiple inpainting steps or a larger inference patch size may be required, with the inpainting speed being loosely proportional to the number of artifacts. 
On the same machine, the 3D GAN approach and GDR (already restricted to the lung bounding box for speed up and reduction of RAM usage) took on average approximately 8\,s and 1\,hour to perform the inpainting.}

\begin{table}
\centering
\scriptsize
\color{black}\caption{Inpainting performance for the external DIR-Lab test data for GDR, GAN, our \textsc{Inpaint}\,[$\mathbf{I}_\mathrm{A}$] and \textsc{Full model} configuration for simulated INT and DS artifacts (performance quantification by $\Delta$RMSE and NCC values; format of the numbers identical to \Cref{tab:results_ablation_study_in-house})}
\label{tab:results_ablation_study_dirlab}
\color{black}\scalebox{0.9}{
\begin{tabular}{wl{0.25cm}wl{0.25cm}wl{0.25cm}*{5}{C{1.1cm}}}
\toprule
\multicolumn{3}{c}{ \# \textbf{slices:}} & \textbf{2} & \textbf{6} & \textbf{10} & \textbf{14} & \textbf{18} \\
\midrule
\multirow{12}{*}{\rotatebox[origin=c]{90}{\textbf{INT artifacts}}} &
\multirow{6}{*}{\rotatebox[origin=c]{90}{$\mathbf{\Delta}$\textbf{RMSE (\%)}}} 
 & GDR$^*$  & --- & --- & --- & --- & --- \\ \addlinespace
&& GAN  & \phantom{$-$}\phantom{0}0.1$_\texttt{- 6.4}^\texttt{~~6.3}$	&	\phantom{$-$}\phantom{0}6.6$_\texttt{~~4.6}^\texttt{~~7.6}$	&	\phantom{$-$}\phantom{0}9.4$_\texttt{~~8.5}^\texttt{~10.5}$	&	\phantom{$-$}\phantom{0}7.4$_\texttt{~~5.4}^\texttt{~~9.1}$	&	\phantom{$-$}10.1$_\texttt{~~7.3}^\texttt{~11.3}$ \\ \addlinespace
&& ours$^{\dagger}$  & \phantom{$-$}\phantom{0}9.1$_\texttt{~~3.5}^\texttt{~10.7}$	&	\phantom{$-$}24.9$_\texttt{~18.7}^\texttt{~26.7}$	&	\phantom{$-$}22.6$_\texttt{~22.1}^\texttt{~25.5}$	&	\phantom{$-$}19.7$_\texttt{~16.8}^\texttt{~20.4}$	&	\phantom{$-$}19.3$_\texttt{~18.2}^\texttt{~19.8}$ \\ \addlinespace
&& ours$^{\ddagger}$  & \bftab\phantom{$-$}25.5$_\texttt{~19.9}^\texttt{~38.6}$	&	\bftab\phantom{$-$}45.2$_\texttt{~38.2}^\texttt{~54.9}$	&	\bftab\phantom{$-$}48.9$_\texttt{~43.3}^\texttt{~62.1}$	&	\bftab\phantom{$-$}54.6$_\texttt{~51.7}^\texttt{~62.7}$	&	\bftab\phantom{$-$}64.5$_\texttt{~51.6}^\texttt{~69.0}$ \\
\cmidrule{2-8}
&\multirow{6}{*}{\rotatebox[origin=c]{90}{\textbf{NCC}}} 
 & GDR$^*$  & --- & --- & --- & --- & --- \\ \addlinespace
&& GAN  &  \phantom{$-$}0.83$_\texttt{~0.81}^\texttt{~0.86}$	&	\phantom{$-$}0.57$_\texttt{~0.52}^\texttt{~0.61}$	&	\phantom{$-$}0.39$_\texttt{~0.38}^\texttt{~0.44}$	&	\phantom{$-$}0.24$_\texttt{~0.23}^\texttt{~0.29}$	&	\phantom{$-$}0.20$_\texttt{~0.20}^\texttt{~0.23}$ \\ \addlinespace
&& ours$^{\dagger}$  & \phantom{$-$}0.85$_\texttt{~0.83}^\texttt{~0.87}$	&	\phantom{$-$}0.60$_\texttt{~0.56}^\texttt{~0.62}$	&	\phantom{$-$}0.45$_\texttt{~0.40}^\texttt{~0.48}$	&	\phantom{$-$}0.29$_\texttt{~0.28}^\texttt{~0.30}$	&	\phantom{$-$}0.15$_\texttt{~0.15}^\texttt{~0.18}$ \\ \addlinespace
&& ours$^{\ddagger}$  & \bftab\phantom{$-$}0.91$_\texttt{~0.90}^\texttt{~0.94}$	&	\bftab\phantom{$-$}0.82$_\texttt{~0.77}^\texttt{~0.89}$	&	\bftab\phantom{$-$}0.80$_\texttt{~0.72}^\texttt{~0.88}$	&	\bftab\phantom{$-$}0.75$_\texttt{~0.73}^\texttt{~0.84}$	&	\bftab\phantom{$-$}0.78$_\texttt{~0.72}^\texttt{~0.83}$ \\
\cmidrule[\heavyrulewidth]{1-8}
\multirow{12}{*}{\rotatebox[origin=c]{90}{\textbf{DS artifacts}}} &
\multirow{6}{*}{\rotatebox[origin=c]{90}{$\mathbf{\Delta}$\textbf{RMSE (\%)}}} 
 & GDR  & \phantom{$-$}\phantom{0}4.1$_\texttt{-10.9}^\texttt{~38.1}$	&	\phantom{$-$}16.9$_\texttt{~10.5}^\texttt{~35.0}$	&	\phantom{$-$}16.1$_\texttt{~~6.9}^\texttt{~31.2}$	&	\phantom{$-$}16.6$_\texttt{~~1.7}^\texttt{~30.6}$	&	\phantom{$-$}18.1$_\texttt{~~5.5}^\texttt{~30.2}$ \\ \addlinespace
&& GAN  & \bftab\phantom{$-$}68.8$_\texttt{~66.8}^\texttt{~70.3}$	&	\phantom{$-$}67.4$_\texttt{~63.3}^\texttt{~71.1}$	&	\phantom{$-$}61.9$_\texttt{~58.2}^\texttt{~67.5}$	&	\phantom{$-$}54.9$_\texttt{~47.7}^\texttt{~63.7}$	&	\phantom{$-$}47.8$_\texttt{~41.6}^\texttt{~60.3}$ \\ \addlinespace
&& ours$^{\dagger}$  & \phantom{$-$}58.2$_\texttt{~54.0}^\texttt{~61.9}$	&	\phantom{$-$}64.8$_\texttt{~61.5}^\texttt{~67.9}$	&	\phantom{$-$}56.8$_\texttt{~53.5}^\texttt{~60.0}$	&	\phantom{$-$}51.1$_\texttt{~46.2}^\texttt{~55.7}$	&	\phantom{$-$}46.7$_\texttt{~42.7}^\texttt{~52.3}$ \\ \addlinespace
&& ours$^{\ddagger}$  & \phantom{$-$}64.2$_\texttt{~60.3}^\texttt{~70.9}$	&	\bftab\phantom{$-$}73.8$_\texttt{~68.1}^\texttt{~79.1}$	&	\bftab\phantom{$-$}70.0$_\texttt{~63.9}^\texttt{~74.6}$	&	\bftab\phantom{$-$}71.3$_\texttt{~56.5}^\texttt{~75.2}$	&	\bftab\phantom{$-$}67.1$_\texttt{~54.6}^\texttt{~72.9}$ \\
\cmidrule{2-8}
&\multirow{6}{*}{\rotatebox[origin=c]{90}{\textbf{NCC}}} 
 & GDR  & \phantom{$-$}0.13$_\texttt{~0.08}^\texttt{~0.27}$	&	\phantom{$-$}0.05$_\texttt{~0.02}^\texttt{~0.10}$	&	\phantom{$-$}0.05$_\texttt{~0.02}^\texttt{~0.11}$	&	\phantom{$-$}0.05$_\texttt{~0.02}^\texttt{~0.09}$	&	\phantom{$-$}0.05$_\texttt{~0.01}^\texttt{~0.10}$ \\ \addlinespace
&& GAN  & \bftab\phantom{$-$}0.77$_\texttt{~0.70}^\texttt{~0.79}$	&	\phantom{$-$}0.51$_\texttt{~0.46}^\texttt{~0.56}$	&	\phantom{$-$}0.36$_\texttt{~0.31}^\texttt{~0.41}$	&	\phantom{$-$}0.22$_\texttt{~0.20}^\texttt{~0.25}$	&	\phantom{$-$}0.18$_\texttt{~0.17}^\texttt{~0.20}$ \\ \addlinespace
&& ours$^{\dagger}$  & \phantom{$-$}0.60$_\texttt{~0.58}^\texttt{~0.64}$	&	\phantom{$-$}0.46$_\texttt{~0.43}^\texttt{~0.49}$	&	\phantom{$-$}0.30$_\texttt{~0.26}^\texttt{~0.33}$	&	\phantom{$-$}0.19$_\texttt{~0.18}^\texttt{~0.23}$	&	\phantom{$-$}0.13$_\texttt{~0.12}^\texttt{~0.15}$ \\ \addlinespace
&& ours$^{\ddagger}$  & \phantom{$-$}0.69$_\texttt{~0.66}^\texttt{~0.83}$	&	\bftab\phantom{$-$}0.63$_\texttt{~0.43}^\texttt{~0.69}$	&	\bftab\phantom{$-$}0.52$_\texttt{~0.35}^\texttt{~0.61}$	&	\bftab\phantom{$-$}0.48$_\texttt{~0.24}^\texttt{~0.61}$	&	\bftab\phantom{$-$}0.49$_\texttt{~0.22}^\texttt{~0.57}$ \\
\bottomrule
\addlinespace
\multicolumn{7}{l}{$^{\dagger}$ \textsc{Inpaint}\,[$\mathbf{I}_\mathrm{A}$] \; $^{\ddagger}$ \textsc{Full model} \; $^*$ inpainting not possible}\\
\end{tabular}}
\end{table}

\subsection{End-to-end validation}
\begin{figure*}
    \centering
    \includegraphics[]{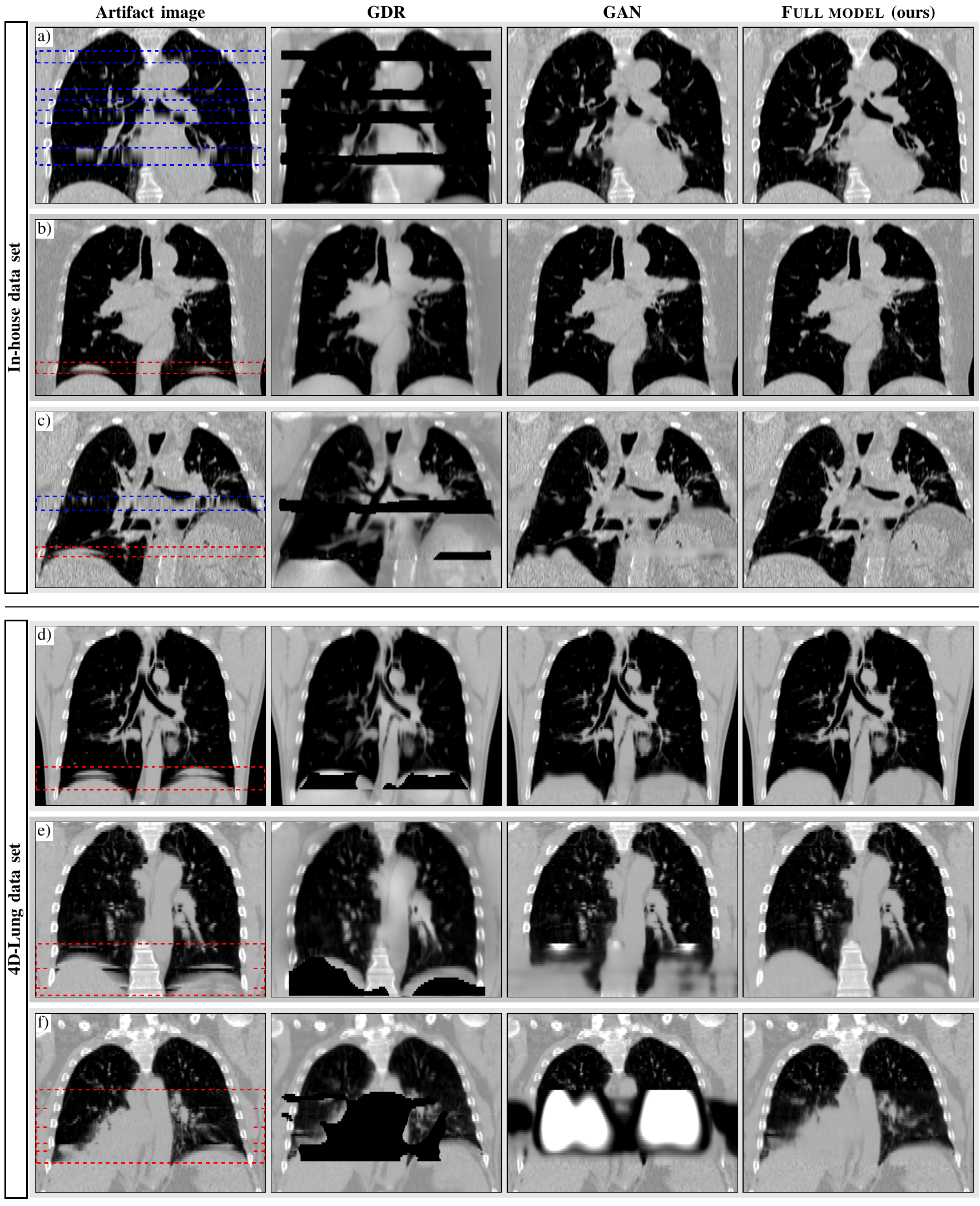}
    \caption{\textcolor{black}{Application of GDR, GAN and our \textsc{Full model} configuration to three in-house (a--c) and three 4D-Lung data sets (d--f) with pronounced INT and/or DS artifacts (cf. \cref{Sec:data}): a) multiple INT, b) small DS, c) small INT and DS, d) small DS, e) multiple clustered DS, f) multiple clustered DS artifacts, occurring in all phase images (potentially an image reconstruction failure). The artifact areas were detected by the artifact detection networks (INT: blue box, DS: red box). The inpainting results for the approaches are shown next to the corresponding artifact image. For example, for the artifact image in a, the inpainted versions are shown in b--d for GDR, GAN and the proposed approach. For GDR, inpainting was restricted to the lung bounding box due to high memory requirements of the algorithm.}
    }\label{fig:inhouse_end_to_end}
\end{figure*}

\textcolor{black}{The natural next step was to combine the developed artifact detection and \textsc{Full model} inpainting configuration to investigate the performance by applying both models in an end-to-end validation scheme on real artifact-affected patient data. Corresponding detection and inpainting results are shown in \cref{fig:inhouse_end_to_end} for six exemplary patients (3 in-house cases, 3 external cases of the 4D-Lung data set) of the artifact data subsets. Additionally and also using the automatically detected artifact to indicate the areas to be inpainted, the GDR and GAN approaches were applied to compare the performance of the proposed to corresponding state-of-the-art methods on real artifact-affected patient data.}

\textcolor{black}{Visual inspection revealed that GDR adequately inpainted only the DS artifact in row\,b of \cref{fig:inhouse_end_to_end}, but failed in all other cases. 
For these cases, a default voxel value ($-$1000, i.e., air HU) was inserted by the algorithm, which, according to the original publication, is the case for insufficient 4D CT image information \cite{shao:2021}. This is particularly noticeable for INT artifacts, for which all phase images of the 4D CT were artifact-affected, and DS artifacts located in areas where intensity-based registration (the basis of GDR) was challenging. Furthermore, the resulting inpainted images appeared always blurred and altered also outside the artifact-affected area(s).} 

\textcolor{black}{As shown in the quantitative evaluation, direct GAN-based inpainting is, to some extent, able to realistically restore image areas with small INT and DS artifacts. It, however, struggles to correctly inpaint artifacts located at the lung/liver border (cf. case c and d in \cref{fig:inhouse_end_to_end}), as well as for inpainting larger artifacts (cf. case e and f in \cref{fig:inhouse_end_to_end}, where the model appeared to be unstable. Similar to our models and due to the underlying design, the GAN model did not alter image areas outside the artifact mask. However, appearance of static objects inside the artifact areas, like rib bones and/or the spine, were considerably changed in comparison to our approach. Similar to the simulated data, our \textsc{Full model} configuration achieves plausible and anatomically realistically appearing results for all cases, even for the clustered DS artifacts in case f of \cref{fig:inhouse_end_to_end}.}

As, different to the simulated artifacts, no ground truth exists for the real image artifacts, no RMSE and NCC could be computed. Instead, we applied the artifact detection networks a second time, now after the inpainting process. \textcolor{black}{On average, the number of artifact-affected axial slices detected after inpainting with our our \textsc{Full model} was reduced by 72\% per data set (INT artifacts: 100\%; DS artifacts: 44\%). 
Due to the total failures visible in, e.g., \cref{fig:inhouse_end_to_end} e and f, corresponding numbers for GDR and GAN could not be computed.}

\begin{table}
\centering
\scriptsize
\color{black}\caption{Inpainting performance for the external 4D-Lung test data for GDR, GAN, our \textsc{Inpaint}\,[$\mathbf{I}_\mathrm{A}$] and the \textsc{Full model} configuration for simulated INT and DS artifacts (performance quantification by $\Delta$RMSE and NCC values; format of the numbers identical to \Cref{tab:results_ablation_study_in-house}).}
\label{tab:results_ablation_study_4d-lung}
\color{black}\scalebox{0.9}{
\begin{tabular}{wl{0.25cm}wl{0.25cm}wl{0.25cm}*{5}{C{1.1cm}}}
\toprule
\multicolumn{3}{c}{ \# \textbf{slices:}} & \textbf{2} & \textbf{6} & \textbf{10} & \textbf{14} & \textbf{18} \\
\midrule
\multirow{12}{*}{\rotatebox[origin=c]{90}{\textbf{INT artifacts}}} &
\multirow{6}{*}{\rotatebox[origin=c]{90}{$\mathbf{\Delta}$\textbf{RMSE (\%)}}} 
 & GDR$^*$  & --- & --- & --- & --- & --- \\ \addlinespace
&& GAN &  \phantom{$-$}14.2$_\texttt{~13.5}^\texttt{~14.9}$	&	$-$\phantom{0}2.5$_\texttt{-14.6}^\texttt{~~9.6}$	&	\phantom{$-$}\phantom{0}3.4$_\texttt{~~2.7}^\texttt{~~4.1}$	&	$-$21.3$_\texttt{-34.5}^\texttt{-~8.1}$	&	$-$19.6$_\texttt{-26.7}^\texttt{-12.5}$ \\ \addlinespace
&& ours$^{\dagger}$  & \bftab\phantom{$-$}52.1$_\texttt{~44.9}^\texttt{~59.8}$	&	\phantom{$-$}\phantom{0}8.8$_\texttt{- 0.4}^\texttt{~35.9}$	&	\phantom{$-$}\phantom{0}9.8$_\texttt{- 4.1}^\texttt{~22.8}$	&	\phantom{$-$}19.6$_\texttt{~16.1}^\texttt{~25.5}$	&	\phantom{$-$}\phantom{0}9.6$_\texttt{~~2.8}^\texttt{~27.0}$ \\ \addlinespace
&& ours$^{\ddagger}$  & \phantom{$-$}39.6$_\texttt{~33.4}^\texttt{~44.1}$	&	\bftab\phantom{$-$}65.4$_\texttt{~58.8}^\texttt{~69.9}$	&	\bftab\phantom{$-$}58.1$_\texttt{~57.3}^\texttt{~63.7}$	&	\bftab\phantom{$-$}66.7$_\texttt{~64.0}^\texttt{~69.9}$	&	\bftab\phantom{$-$}33.9$_\texttt{~26.3}^\texttt{~52.9}$ \\
\cmidrule{2-8}
&\multirow{6}{*}{\rotatebox[origin=c]{90}{\textbf{NCC}}} 
 & GDR$^*$  & --- & --- & --- & --- & --- \\ \addlinespace
&& GAN  & \phantom{$-$}0.49$_\texttt{~0.47}^\texttt{~0.50}$	&	\phantom{$-$}0.22$_\texttt{~0.18}^\texttt{~0.25}$	&	\phantom{$-$}0.18$_\texttt{~0.17}^\texttt{~0.18}$	&	\phantom{$-$}0.14$_\texttt{~0.13}^\texttt{~0.15}$	&	\phantom{$-$}0.06$_\texttt{~0.05}^\texttt{~0.08}$ \\ \addlinespace
&& ours$^{\dagger}$  &	\phantom{$-$}0.56$_\texttt{~0.53}^\texttt{~0.59}$	&	\phantom{$-$}0.27$_\texttt{~0.21}^\texttt{~0.32}$	&	\phantom{$-$}0.19$_\texttt{~0.15}^\texttt{~0.22}$	&	\phantom{$-$}0.12$_\texttt{~0.06}^\texttt{~0.15}$	&	\phantom{$-$}0.07$_\texttt{~0.04}^\texttt{~0.09}$ \\ \addlinespace
&& ours$^{\ddagger}$  &	\bftab\phantom{$-$}0.75$_\texttt{~0.69}^\texttt{~0.75}$	&	\bftab\phantom{$-$}0.72$_\texttt{~0.70}^\texttt{~0.78}$	&	\bftab\phantom{$-$}0.65$_\texttt{~0.65}^\texttt{~0.70}$	&	\bftab\phantom{$-$}0.65$_\texttt{~0.65}^\texttt{~0.71}$	&	\bftab\phantom{$-$}0.67$_\texttt{~0.65}^\texttt{~0.71}$ \\
\cmidrule[\heavyrulewidth]{1-8}
\multirow{12}{*}{\rotatebox[origin=c]{90}{\textbf{DS artifacts}}} &
\multirow{6}{*}{\rotatebox[origin=c]{90}{$\mathbf{\Delta}$\textbf{RMSE (\%)}}} 
& GDR & \phantom{$-$}\phantom{0}1.4$_\texttt{-49.0}^\texttt{~25.7}$	&	\phantom{$-$}45.1$_\texttt{- 6.3}^\texttt{~54.0}$	&	\phantom{$-$}26.1$_\texttt{- 6.0}^\texttt{~27.9}$	&	\phantom{$-$}22.0$_\texttt{- 5.2}^\texttt{~23.2}$	&	\phantom{$-$}18.6$_\texttt{-12.4}^\texttt{~19.9}$ \\ \addlinespace
&& GAN & \phantom{$-$}46.6$_\texttt{~46.2}^\texttt{~47.1}$	&	\phantom{$-$}27.4$_\texttt{~26.7}^\texttt{~28.1}$	&	\phantom{$-$}16.2$_\texttt{~11.1}^\texttt{~21.4}$	&	\phantom{$-$}16.0$_\texttt{~14.5}^\texttt{~17.5}$	&	\phantom{$-$}\phantom{0}9.4$_\texttt{~~6.5}^\texttt{~12.3}$ \\ \addlinespace
&& ours$^{\dagger}$ & \phantom{$-$}46.7$_\texttt{~46.3}^\texttt{~47.1}$	&	\phantom{$-$}35.8$_\texttt{~35.4}^\texttt{~36.3}$	&	\phantom{$-$}25.8$_\texttt{~19.8}^\texttt{~31.8}$	&	\phantom{$-$}32.4$_\texttt{~29.7}^\texttt{~35.1}$	&	\bftab\phantom{$-$}35.4$_\texttt{~32.6}^\texttt{~38.2}$ \\ \addlinespace
&& ours$^{\ddagger}$ & \bftab\phantom{$-$}48.0$_\texttt{~42.7}^\texttt{~52.1}$	&	\bftab\phantom{$-$}61.5$_\texttt{~57.2}^\texttt{~63.4}$	&	\bftab\phantom{$-$}50.7$_\texttt{~50.2}^\texttt{~56.4}$	&	\bftab\phantom{$-$}43.9$_\texttt{~43.3}^\texttt{~51.9}$	&	\phantom{$-$}33.8$_\texttt{~33.2}^\texttt{~43.5}$ \\
\cmidrule{2-8}
&\multirow{6}{*}{\rotatebox[origin=c]{90}{\textbf{NCC}}} 
& GDR &	\phantom{$-$}0.11$_\texttt{~0.05}^\texttt{~0.21}$	&	\phantom{$-$}0.17$_\texttt{~0.09}^\texttt{~0.34}$	&	\phantom{$-$}0.06$_\texttt{~0.03}^\texttt{~0.07}$	&	\phantom{$-$}0.05$_\texttt{~0.03}^\texttt{~0.06}$	&	\phantom{$-$}0.05$_\texttt{~0.02}^\texttt{~0.05}$ \\ \addlinespace
&& GAN &	\phantom{$-$}0.33$_\texttt{~0.31}^\texttt{~0.35}$	&	\phantom{$-$}0.16$_\texttt{~0.16}^\texttt{~0.16}$	&	\phantom{$-$}0.08$_\texttt{~0.07}^\texttt{~0.08}$	&	\phantom{$-$}0.09$_\texttt{~0.08}^\texttt{~0.09}$	&	\phantom{$-$}0.09$_\texttt{~0.08}^\texttt{~0.09}$ \\ \addlinespace
&& ours$^{\dagger}$ &	\phantom{$-$}0.34$_\texttt{~0.32}^\texttt{~0.35}$	&	\phantom{$-$}0.18$_\texttt{~0.18}^\texttt{~0.19}$	&	\phantom{$-$}0.12$_\texttt{~0.11}^\texttt{~0.12}$	&	\phantom{$-$}0.12$_\texttt{~0.12}^\texttt{~0.13}$	&	\phantom{$-$}0.11$_\texttt{~0.11}^\texttt{~0.11}$ \\ \addlinespace
&& ours$^{\ddagger}$ &	\bftab\phantom{$-$}0.56$_\texttt{~0.53}^\texttt{~0.58}$	&	\bftab\phantom{$-$}0.51$_\texttt{~0.47}^\texttt{~0.52}$	&	\bftab\phantom{$-$}0.41$_\texttt{~0.33}^\texttt{~0.42}$	&	\bftab\phantom{$-$}0.34$_\texttt{~0.28}^\texttt{~0.38}$	&	\bftab\phantom{$-$}0.25$_\texttt{~0.21}^\texttt{~0.30}$ \\
\bottomrule
\addlinespace
\multicolumn{7}{l}{$^{\dagger}$ \textsc{Inpaint}\,[$\mathbf{I}_\mathrm{A}$] \; $^{\ddagger}$ \textsc{Full model} \; $^*$ inpainting not possible} \\
\end{tabular}}
\end{table}

% ---------- DISCUSSION SECTION:
\section{Discussion}

We presented a deep learning-based conditional inpainting approach for 4D CT data that automatically detects and restores image areas affected by 4D CT motion artifacts. 
\textcolor{black}{In contrast to recent publications on 4D CT artifact reduction and inpainting methods \cite{shao:2021,jin:2021}, the proposed system is able to robustly detect and correct both common 4D CT artifact types (double structure and interpolation artifacts) within seconds. Even for out-of-distribution artifact sizes, as seen for 4D-Lung cases e and f in \cref{fig:inhouse_end_to_end}, our approach is able to generate an inpainted version of the initial artifact-affected image that appears anatomically plausible.}

Key to the performance of the developed system were the following methodical aspects and contributions:

\noindent
\textbf{Conditional image:} Integrating prior knowledge about the patient-specific anatomy, routinely acquired in clinical practice, into the inpainting process improved the performance. This was particularly evident with larger artifacts \textcolor{black}{when compared to direct inpainting without prior knowledge (GAN, \textsc{Inpaint}\,[$\mathbf{I}_\mathrm{A}$])}.

\noindent
\textbf{3D input data:} Straightforward application of 2D inpainting networks available for the natural image domain is not sufficient for 4D CT data. To take into account out-of-plane motion of anatomical structures, 3D inpainting approaches were needed (here: a patch-based approach particularly designed to ensure fast but still accurate inpainting results).

\noindent
\textbf{Partial convolutions:} Inpainting using standard convolutional layers failed in particular for INT artifacts, where the conditional image offers only a blurred representation of the patient's anatomy. The restriction of the convolution to valid voxels only and the successive filling of the artifact area, i.e., partial convolutions, helped to improve the model stability during training as well as the resulting inpainting performance.

\noindent
\textbf{Spatial transformers:} The position of the anatomical structures in the conditional images do not necessarily match the positions in the artifact-affected images -- with the discrepancies increasing for large motion amplitudes and artifact images close to the end-inspiration breathing phase. Integrating an end-to-end trainable deformable variant of a spatial transformer block to align the anatomical information of the conditional and the artifact-affected image before inpainting helped to reduce related uncertainties. 

Furthermore, by integrating straight-forward physiological prior knowledge of the effects that cause 4D CT motion artifacts, we were able to realistically simulate artifacts and superimpose such artifacts on artifact-free image data. Pairs of artifact-free ground truth and artifact-affected images were generated and enabled quantitative evaluation of the proposed system by standard image similarity metrics between artifact-free and artifact-affected data before/after inpainting. \textcolor{black}{Therefore, even the publicly available, almost artifact-free 4D CT data of the DIR-Lab and the 4D-Lung repositories could be used for analyzing the inpainting performance and to illustrate generalizability of the proposed approach to 4D CT data acquired at different centers and CT scanners.}

From an application perspective, short inference times of the artifact detection and subsequent artifact inpainting\textcolor{black}{, especially compared to registration-based approaches like GDR,} indicate feasibility of direct integration into clinical workflows. For instance, after acquisition and reconstruction of a 4D CT, the DL-based conditional inpainting approach could be employed to prevent a potentially necessary re-scan due to motion artifacts. This is line with the ALARA (as low as reasonably achievable) principle of radioprotection and at the same time allows for an accurate radiotherapy planning process (e.g., reliable contouring of organs at risk/target volumes, dose calculation). The latter aspect will even be more relevant for 4DCT-ventilation-based radiation therapy treatment planning \cite{Vinogradskiy2021}. 

Nevertheless, the present study contains some limitations to be addressed. Using image information of a conditional image for inpainting requires the conditional image to be more or less artifact-free. 
Here, this was achieved by choosing conditional images that are usually not affected by artifacts: the end exhalation phase for DS artifacts and the temporal average CT for INT inpainting. 
However, it is in principle possible that the end exhalation phase image is also affected by DS artifacts. In this case, the inpainting performance will decrease. 
This could also be an explanation for the observation that even after inpainting, the artifact detection network identifies 56\% of the original DS artifact areas as still artifact-affected, although the artifacts are visually clearly reduced and hardly visible.  
In turn, for INT artifacts, the temporal average CT becomes increasingly blurred with larger motion amplitudes. Thus, fine structural details can be lost during inpainting. 
\textcolor{black}{However, the blurred appearance of the inpainted structures also becomes visible for larger DS artifacts and could  generally be a consequence of a larger uncertainty of the model for larger areas to be inpainted. The blurred structure appearance is, nevertheless, for the proposed approach less pronounced than for GDR and standard GAN-based inpainting (cf. \cref{fig:inhouse_end_to_end}).}

\textcolor{black}{Despite these limitations,} the results demonstrate the potential of the proposed approach to restore artifact-affected 4D CT images.
Furthermore, the presented concept could also be transferred to operate on non motion-related artifacts (e.g., metal, beam hardening or streaking artifacts)
\textcolor{black}{and other anatomical areas and organs. However, this requires retraining of the networks to adapt to the specific artifacts and image characteristics}.  

\section{Conclusions}
The proposed DL-based conditional artifact inpainting approach and the corresponding results highlight the promising potential of current deep learning architectures and techniques for correcting  typical 4D CT artifacts -- and, thereby, to reduce uncertainties during 4D radiation therapy treatment planning. 

\bibliographystyle{ieee}
\bibliography{library}
\end{document}